\documentclass[11pt]{article} 
\usepackage{epsfig, graphicx}
\usepackage[multiple]{footmisc}
\usepackage{color,cite}
\usepackage{hyperref, url}
\usepackage[textwidth=7.5in,textheight=9.7in]{geometry}

\def\eqn#1{eq.\ (\ref{#1})}

\def\bea{\begin{eqnarray}}
\def\eea{\end{eqnarray}}

\begin{document}
\title{\bf New class of naked singularities and their observational signatures}
\author{Kaushik Bhattacharya$^a$\footnote{e-mail: \tt kaushikb@iitk.ac.in},
Dipanjan Dey$^a$\footnote{e-mail: \tt deydip@iitk.ac.in}, 
Arindam Mazumdar$^b$\footnote{e-mail: \tt arindam.mazumdar@iitkgp.ac.in}
, Tapobrata Sarkar$^a$\footnote{e-mail: \tt tapo@iitk.ac.in}
}
\date{}
\maketitle
\vspace*{-1cm}
\begin{center}
{‌\small $a$) Department of Physics, Indian Institute of Technology, Kanpur-208016, India \\
$b$) Centre for Theoretical Studies, Indian Institute of Technology, Kharagpur, West Bengal-721302, India} 
\end{center}
 
\begin{abstract}
{By imposing suitable junction conditions on a space-like hyper-surface, we obtain a two-parameter family of possible static
configurations from gravitational collapse. These exemplify a new class of naked singularities. We show that 
these admit a consistent description via a two-fluid model, one of which might be dust. We then study lensing
and accretion disk properties of our solution and point out possible differences with black hole scenarios.
The distinctive features of our solution, compared to the existing naked singularity solutions in the literature
are discussed.}
\end{abstract}
\maketitle
\section{Introduction}\label{introduction}

The study of gravitational collapse within the framework of General Relativity (GR) \cite{Weinberg},\cite{Poisson} 
is a well researched topic that continues to attract a lot of attention. Tracking a system collapsing 
under its own gravity is indeed a formidable task and seminal papers on the topic have appeared in the past several 
decades (see, e.g \cite{Oppenheimer:1939ue, Penrose,Penrose2,Hawking,Misner:1964je,May:1966zz,Hawking2,Choptuik}). 
In typical collapse scenarios, an internal collapsing metric is matched across a time-like hyper-surface to an
external space-time via the junction conditions \cite{Israel}. 

It is not difficult to envisage a collapse process in which such junction conditions are applied to 
a space-like hyper-surface instead. Indeed, while conservation laws prevent the formation of a matter
shell on a space-like hyper-surface, smooth matching (without matter) across such a hyper-surface is 
a justifiable process in GR (see the discussion in section 21.13 of Misner, Thorne and Wheeler\cite{MTW}). 
To make the picture more concrete, we can visualize a possible primordial over-dense patch of dust, described by
a closed Friedman-Lemaitre-Robertson-Walker (FLRW) metric,
which initially undergoes expansion along with a background FLRW solution. Then, at some turn around time, it
decouples from the background and starts collapsing, with the collapse process described by a different
space-time metric. To make this evolution smooth, we match the two metrics 
on a space-like hyper-surface at a certain co-moving time, which, for simplicity, we choose as the
turn around time. Upon decoupling, the collapsing over-dense region is matched smoothly to a Schwarzschild
metric, which can occur if the radial pressure is zero, an assumption that we use here (see below). 
The Schwarzschild metric in turn matched with the expanding FLRW dust background.

We then look for possible (meta)stable configurations that arise in such a collapse process 
with non-trivial restrictions imposed by the junction conditions. These are configurations in which the 
velocity and acceleration of the collapsing shells vanish simultaneously. It is well known \cite{Markovic}
that such configurations are possible to attain in collapse with a cosmological constant. 
In scenarios without a cosmological constant, such a situation, while impossible to attain for dust collapse, 
can occur in examples with pressure \cite{Joshi:2011zm} with a central naked singularity. 
Here, as we have mentioned, we will choose a collapse configuration with vanishing radial pressure but non-vanishing 
tangential pressures, as a simple model. Such spherically symmetric configurations
supported only by tangential stresses is well known. In the static case, these were originally considered by
Einstein \cite{EinsteinCluster} and are called Einstein clusters. An interior Schwarzschild solution with vanishing
radial pressure was obtained by Florides in \cite{Florides}, and are called the Florides solution, 
which under certain conditions, may be identified with the Einstein cluster solution.\footnote{Studies on the possibility of 
modeling galactic halos as Einstein clusters have appeared in the works of Lake \cite{LakeCluster} and Boehmer and 
Harko \cite{BoehmerHarko}. }
Non-static generalizations of this model has been considered by 
Datta \cite{Datta} and Bondi \cite{Bondi}. Collapse in models with vanishing radial stress has been originally considered
by Magli in \cite{Magli1}, \cite{Magli2}. More recently, meta-stable static configurations discussed above in 
with vanishing radial pressure were presented by Joshi, Malafarina and Narayan (JMN) in \cite{Joshi:2011zm}. 
These are dubbed in the literature as the JMN naked singularity solutions. 

In this work, we find a new two-parameter family of naked
singularities that can arise out of the collapse process that we have just discussed, and is very different in 
nature than the one-parameter solution obtained by JMN. As a special case, we recover the interior Schwarzschild solution 
of Florides, so in some sense our solution is a generalization of the latter, as is the JMN solution.
Note that our collapse process here is distinct from the conventional structure formation scenario,
via the process of virialization \cite{KolbTurner}, and should be taken as an idealized toy model that produces
a class of naked singularities as its end stage. We can nonetheless make some
estimates of observable quantities from our simple model, by choosing typical realistic values of the central mass.
In this spirit, we analyse these new solutions further, and establish some 
distinguishable properties of gravitational lensing by modeling the central mass as a $10^7M_{\odot}$ object and contrasting
its features with a black hole of similar mass.  
We also study accretion disk properties of our solution. These are shown to be distinctly
different from black hole backgrounds and the JMN naked singularity backgrounds. 
We further consider the nature of the fluid sourcing our space-time. 
This is an important issue, as the naked singularity backgrounds known in the literature do not have
a simple equation of state (note that the Florides solution or the interior Schwarzschild solution due to Synge \cite{Florides}
are constant density solutions and do not have an obvious equation of state). 
We show here that although the latter fact continues to be true, we can 
write the energy momentum tensor of our solution in terms of two non-interacting perfect fluids, one of
which can be dust. Some general properties of such a picture in more generic backgrounds with 
zero radial pressure is also commented upon. 

All of the above assumes importance in the context of the results from the Event Horizon Telescope (see
\cite{EHT1} and the followup papers). Indeed a major thrust of recent research in GR is to understand 
images and shadows formed by singularities, and it is important to distinguish and quantify observational signatures 
from black holes and other singular objects such as naked singularities, as the latter can often mimic black holes. 
Recently, there has been an upsurge in the literature on the possible observational distinctions between black holes and 
naked singularities, as a complete proof of the cosmic censorship conjecture is still lacking (for a small sampling
of the recent literature, see, e.g \cite{NSO1, NSO2, NSO3, NSO4, NSO5, NSO6}). 

Popular models of naked singularities that have been well studied in the literature are the over-spinning Kerr solution, 
the over-charged Reissner-Nordstrom solution, the Janis-Newman-Winicour (JNW) space-time and its rotating 
generalization, the JMN solution, etc. Lensing properties of naked singularities was first studied in \cite{VENaked} for the JNW solution, 
and later extended to its rotating version in \cite{GyulNaked}. Lensing in the JMN background has recently
been studied in \cite{Rajibul1}, \cite{Rajibul2}. Accretion disk properties of naked singularities 
have also been studied extensively (see, e.g \cite{Kovacs:2010xm}), and tidal forces in naked singularity
backgrounds have been studied in \cite{TSone}, \cite{TStwo}. 
Our results further such investigations with a novel background naked singularity model. Importantly, as we show here,
observational predictions of our model are very different from the ones that exist in the literature. For example, a study
of null geodesics in our model establishes the existence of stable light rings (anti-photon spheres) which is notably
absent in the JMN space-time and hence the lensing properties of our solution is very different from the ones
in the JMN background. Further, whereas the accretion disk properties of the JMN space-time dictates that 
the radiative efficiency of this model is always $100\%$, that of our model can vary, depending on the 
choice of parameters. These facts 
highlight the importance of our results, and put them in recent perspective. 

This work is organized as follows. In section~\ref{gencol}, we briefly describe the collapse 
process with a generic collapsing metric and formulate the junction conditions and briefly
discuss the initial data. 
In section~\ref{aniso} we take a concrete model to first comment upon the relevant energy conditions, 
show the effects of the constraints derived in the section~\ref{gencol}, and then proceed to derive 
a probable end-state configuration. Next in this section, we proceed to study the nature of the fluid sourcing the 
singularity, and show that
this has a very simple interpretation as a two-fluid model, one whose components is dust. 
Important observational aspects of this 
end state is described in section~\ref{physical}, where we first discuss strong gravitational lensing, and 
later, the accretion disk properties of
our solution. We end with a summary of our main results in section~\ref{concl}.
  
\section{Formulation, junction conditions, and the initial data}
\label{gencol}

The scenario that we will study here is as follows. We assume that an initially over-dense
region, described by a closed FLRW metric undergoes expansion up to a turnaround time, along
with a background FLRW region. After this time, the over-dense region decouples from the background and
starts collapsing. The collapsing
space-time that has completely decoupled from the initial FLRW patch 
may induce inhomogeneous fluid energy-density or may even
have pressure of the collapsing fluid. The initial over density
expands in a homogeneous and isotropic fashion carrying the traits of
the background expansion. Once the turnaround time has reached and the
over dense region collapses, this latter collapsing phase does not bear any
resemblance to the background expansion, and it need not carry the traits of
the background expansion any more. Thus the collapsing sub universe in
principle can have inhomogeneous matter. 

We will set up a situation here where
the initial data for the collapse is obtained by matching the spatially closed FLRW metric
with the general collapsing metric on a space-like hypersurface via the junction conditions \cite{Israel}. 
We will work in a scenario where the Minser-Sharp mass inside the collapsing matter is fixed,
so that the collapsing metric is always matched with an external Schwarzschild one, at 
a time-dependent radius. 
Under suitable conditions, at large co-moving time, the collapse will halt, i.e the matter shells will not fall into a
singularity, but rather produce a static meta-stable state, while matched with an 
external Schwarzschild space-time. This meta-stable state nonetheless has a central singularity which will
be visible to an external Schwarzschild observer, and thus will exemplify a naked singularity. 

\begin{figure}[h!]
\centering
 \includegraphics[width=6.00in]{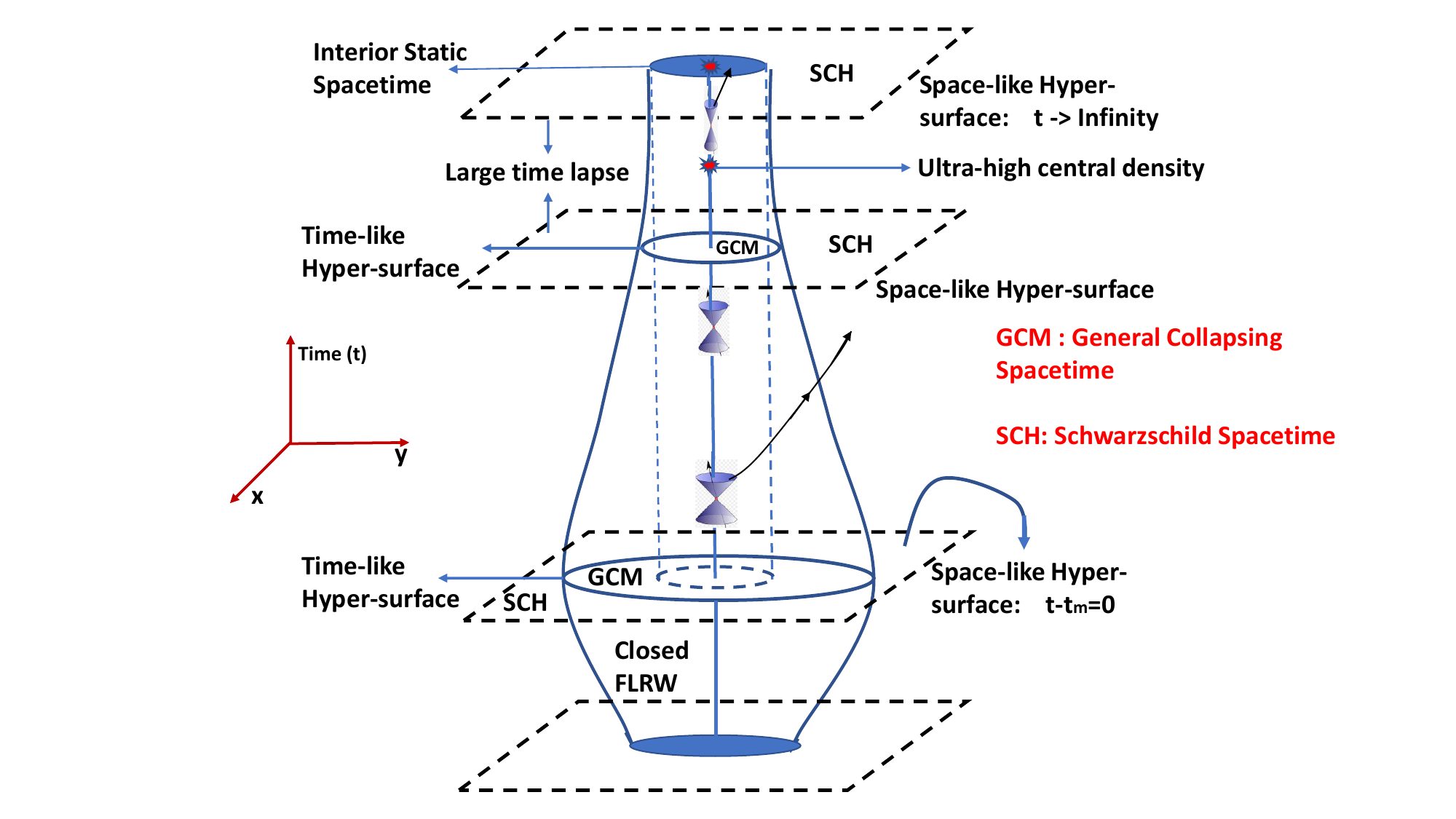}
 \caption{Space-time diagram of the collapse process.}
 \label{std}
\end{figure}
A space-time diagram of the collapse process is shown in fig.(\ref{std}). 
At the space-like hypersurface $t - t_{\rm max}=0$, we match a closed FLRW universe with
a general collapsing metric. 
This then undergoes collapse as determined by the initial data, until the collapse
halts at large times, producing a static configuration with a divergent central density.

\subsection{Metric and junction conditions}

The metric for the two regions, following our previous remarks, are given by
\begin{eqnarray}
ds_-^2 &=&- dt^2 + \frac{a^2(t)}{1- r^2 }dr^2 + a^2(t) r^2 d\Omega^2~\nonumber\\
ds_+^2 &=& - e^{2\nu(r,t)} dt^2 + {R'^2\over G(r,t)}dr^2 + R^2(r,t) d\Omega^2~,
\label{both}
\end{eqnarray}
and we will generally work in natural 
units with the Newton's constant and the speed of light set to unity. 
Here, $\nu(r,t)$, $R(r,t)$ and $G(r,t)$ that appear in the collapsing metric, are functions of
the co-moving radius and time, i.e, $r$ and $t$ respectively, and $d\Omega^2$ is the metric on the unit $2$-sphere.
The range of the coordinates describing the contracting
space-time, are assumed to remain the same as that of 
the expanding phase. Here and in general, a prime (dot) above a function specifies
the derivative of that function with respect to the radial (time) coordinate. Note that for a collapsing
scenario ${\dot R}(r,t) < 0$ at all co-moving times. 

Let us first collect the relevant details of the space-time described by $ds_+^2$ above. We will write \cite{Weinberg}
\begin{eqnarray}
R(r,t)= r f(r,t)\,,
\label{scaling}
\end{eqnarray}
where the functional form of $f(r,t)$ can be known at the time when
collapse starts if one knows the value of $R(r,t)$ at that time. We also define a
function $M(r,t)$, called the Misner-Sharp mass \cite{May:1966zz}, that specifies the amount
of matter enclosed by a shell labeled by $r$, and its functional
form is given by
\begin{eqnarray}
M = R\left(1 - G + e^{-2\nu} \dot{R}^2\right)\,.
\label{ms}  
\end{eqnarray}
In the case of a fluid whose energy-momentum tensor components are given as
$T^0_{\,\,\,\,0}=-\rho$, $T^1_{\,\,\,\,1}=P_r$,
$T^2_{\,\,\,\,2}=T^3_{\,\,\,\,3}=P_\perp$, the Einstein's equations give, in terms of $M(r,t)$,
\begin{eqnarray}
\rho=\frac{M^\prime}{R^2 R^\prime}\,,\,\,\,\,
P_r = -\frac{\dot{M}}{R^2 \dot{R}}\,,\,\,\,\,
P_\perp = \frac12 \rho R \frac{\nu^\prime}{R^\prime}\,,
\label{eineqns}
\end{eqnarray}
where the last relation is valid when the radial pressure $P_r$ vanishes, which is the case that we will be
primarily interested in, and we have denoted the tangential stresses $P_{\theta} = P_{\phi} \equiv P_{\perp}$.
Also, the vanishing of the non-diagonal terms in the Einstein
equations produces a fourth equation :
\begin{eqnarray}
\dot{G}=2\frac{\nu^\prime}{R^\prime}\dot{R}G\,.
\label{vanond}
\end{eqnarray}

To understand gravitational collapse in the scenario described by \eqn{both}, we need to know the
form of the functions $\nu(r,t)$, $R(r,t)$ and $G(r,t)$ at the onset
of collapse. In general it might be difficult to get these initial
data about the functions. Since we are using a scenario in which one
retains the initial expanding phase up to a turn-around point, i.e, when $t=t_{\rm max}$ and $a=a_{\rm
max}$ (with $a_{\rm max}$ assumed to be finite), and then match the spatially closed FLRW metric with the
general collapsing metric to describe
the final collapsing phase, the functions $\nu(r,t)$,
$R(r,t)$ and $G(r,t)$ can be evaluated at $t=t_{\rm max}$ by
junction conditions (see, e.g \cite{Poisson}) applied on a space-like hyper-surface (see fig.(\ref{std})). 

For smooth matching on the space-like hyper-surface $\phi=
t-t_{\rm max}=0$, we first match the induced metrics on both side of
this hyper-surface. This is seen to imply that the initial values of $\nu(r,t)$, $f(r,t)$ and 
$G(r,t)$ at $t_{\rm max}$ should be
\begin{eqnarray}
\nu(r,t_{\rm max})=0,~~f(r,t_{\rm max}) = a_{\rm max},~~ G(r,t_{\rm max}) = 1-r^2\, .
\label{nmatch}
\end{eqnarray}
It can be checked that the extrinsic curvature components arising on a space-like hyper-surface
from the FLRW metric are all proportional to ${\dot a}$ and thus vanish at $t=t_{\rm max}$. 
Hence, we will also require that the extrinsic curvatures due to the metric for $ds_+^2$ of 
\eqn{both} should vanish at this matching surface. To
glean insight into this set of constraint, we first record the
expressions for the components,
\begin{eqnarray}
K_{rr} &=& \frac{e^{-\nu (r,t)}}{2 G(r,t)^2}\left(r f'(r,t)+f(r,t)\right)\left(\left(r
   f'(r,t)+f(r,t)\right) {\dot G}(r,t)-2 \left({\dot f}(r,t)+r{\dot f}'(r,t)\right) G(r,t)\right)\nonumber\\
K_{\theta\theta} &=& -r^2 f(r,t) {\dot f(r,t)} \left(e^{-\nu (r,t)}\right) = \frac{K_{\phi\phi}}{\sin^2\theta}~.
\end{eqnarray}
It is thus seen that vanishing of the
extrinsic curvatures $K_{\theta\theta}$ and $K_{\phi\phi}$ at $t=t_{\rm max}$ requires $\dot f(r, t_{\rm max}) =0$,
and since $f(r,t)$ is independent of $r$ at $t=t_{\rm max}$, we further require
$\dot{G}(r,t_{\rm max})=0$ for vanishing of $K_{rr}$. The physical interpretation of the last two conditions is clear.
Namely, at $t=t_{\rm max}$, the expansion momentarily halts and collapse begins, so that ${\dot R}(r, t_{\rm max}) = 0$. From 
eq.(\ref{scaling}), this implies that $\dot f(r, t_{\rm max}) =0$, with $\dot{G}(r,t_{\rm max})=0$ then following
from eq.(\ref{vanond}). Importantly, we have determined the form of the Misner-Sharp mass of eq.(\ref{ms}).
In our case $P_r=0$ translates to 
$\dot{M}=0$, and consequently the form of
$M$ can be fixed if we know it at $t=t_{\rm max}$. From the matching conditions of \eqn{nmatch}, we get
\begin{eqnarray}
M(r) \equiv M(r,t_{\rm max}) = rf(r,t_{\rm max})\left[1- G(r,t_{\rm max}) +
r^2 e^{-2\nu(r,t_{\rm max})} \dot{f}^2(r,t_{\rm max})\right]= a_{\rm max} r^3\,,
\label{msm}
\end{eqnarray}
which is a regular function throughout the contracting phase.

\subsection{The initial data}

The discussion in this subsection follows \cite{Joshi:2011zm}, \cite{JMN2}. 
Since the Misner-Sharp mass is independent of time, the collapsing metric is always matched with 
an external Schwarzschild solution (as indicated in fig.(\ref{std})). From the Einstein's equations of
eq.(\ref{eineqns}), we then obtain
\begin{equation}
\rho(r,t_{max}) = \frac{3}{a_{\rm max}^2}~,~~P_{\perp}(r,t_{\rm max}) = 0~,
\label{initialdata}
\end{equation}
where we have used eq.(\ref{nmatch}). Thus we see that the initial data is regular at the origin, since 
$a_{\rm max}$ is assumed to be finite. We thus have gravitational collapse starting from regular 
initial data. Now, in order for the collapse model to be physically reasonable, one needs to impose
a few conditions \cite{Joshibook}, \cite{JMN2}. 
First of all, we will assume that there are no shell crossing singularities, which is 
equivalent to the condition $R'>0$ \cite{JoshiSaraykar}. Under this assumption, the density $\rho$ of
eq.(\ref{eineqns}) is positive definite. Now if $\nu' >0$, then the weak and strong
energy conditions will be satisfied. Although it is difficult to justify this assumption all along the collapse,
we will show in sequel that close to the formation of the final singularity, this condition can be expected
to be true. Finally, we will require the pressure gradient and the density to remain regular at the origin before the
formation of the singularity. It can be checked that this implies in our case, $f(0,t) \neq 0$ from the time the
collapse starts, till the time the singularity forms. As we will see later, $f(0,t) = 0$ at large co-moving times, i.e
when the singularity forms. 

It is also to be noted that while the tangential pressure of the collapsing fluid, which is zero at $t=t_{\rm max}$ 
is a non-trivial function of the co-moving time as the collapse progresses. Again, in the absence of an analytic
solution throughout the collapse process, it is difficult to make a concrete statement regarding its behavior 
at all times. However, we can reasonably assume that its time derivative (which can be obtained from 
eq.(\ref{eineqns})) is small, i.e it gradually evolves to its equilibrium value. A similar assumption will be made 
for the density evolution as well.

At the equilibrium that results at the end state of collapse (to be elaborated upon shortly), we can write
\begin{equation}
\rho_e = \frac{3a_{\rm max}}{f_e^3 + rf_e^2f_e'}~,
\label{densityeq}
\end{equation}
where $e$ denotes the value at equilibrium, where $\rho_e$ and $f_e$ depend only on the radial coordinate. 
If $f_e$ is a constant, then we recover a constant density solution at the origin, but if 
$f_e(0) = 0$, then we have a central singularity which has resulted from the collapse process with regular 
initial data. If for simplicity we choose a regular power law behavior 
$f_e \sim r^{\alpha}$, the above discussion implies that we have to choose a non-negative
value of $\alpha$. 

\section{Collapse with anisotropic pressure}\label{aniso}

We now study the collapse of the system described in the previous section, subsequent to $t_{\rm max}$. 
The dynamical evolution equations are given in eqs.(\ref{eineqns}) and (\ref{vanond}). We will be 
interested in the end stage of collapse. As we will see subsequently, it is possible to reach a static
meta-stable configuration with a central singularity in infinite coordinate time. 

\subsection{Dynamics and energy conditions}

The system of equations as given in
\eqn{eineqns} and \eqn{vanond} can produce interesting
final configurations of the space-time if we know about
only two functions. Specifically, once we set $P_r=0$, we have six
different functions as $\nu$, $G$, $M$, $P_\perp$, $\rho$ and $f$ at
our hand and there are four equations in \eqn{eineqns} and
\eqn{vanond}. Since the Misner-Sharp mass has already been defined, we have to now specify the form
of one more function to close the equations, and we discuss below that a good choice is the function $f(r,t)$ at equilibrium. 
Note that we have not specified an equation of state here, which could have been another viable approach. 
Indeed, as pointed out in \cite{JMN2}, there is no compelling reason to choose a fixed equation of state as collapse 
proceeds. In our case, this is determined as an outcome of the final equilibrium configuration. 

This configuration is reached when collapse halts, and the velocity and acceleration of the collapsing shells vanish
simultaneously. Then, we have to satisfy the conditions 
\begin{eqnarray}
{\dot R} = {\ddot R} = 0~,~ {\rm i.e}~\dot{f} = \ddot{f} =0\,,
\label{stabc}
\end{eqnarray}
where in writing the second identity, we have used eq.(\ref{scaling}). This set of conditions is more conveniently written in terms of an
effective potential with independent variables $r$ and $f(r,t)$. 
This is clear from eq.(\ref{ms}), from where we see that the equation of motion for $R$ can be written as 
\begin{equation}\label{eqom}
{\dot R}^2 = e^{2\nu}\left(G - 1 + \frac{M}{R}\right)~,~~{\rm i.e} ~{\dot f}^2 = e^{2\nu}\left(\frac{G - 1}{r^2} + \frac{M}{r^3f}\right)~
= -V(r,f)~,
\end{equation}
with $V(r,f)$ being an effective potential. Then, ${\dot f}=0$ implies that $V=0$ and 
${\ddot f}=0$ can be seen to imply $V_{,f}=0$. 
From eq.(\ref{eqom}), it is then seen that this second condition implies
\begin{equation}\label{ddot}
e^{2\nu}\left[\left(\frac{G_{,f}}{r^2} - \frac{M}{r^3f^2}\right) + 2\nu_{,f}\left(\frac{G - 1}{r^2} + \frac{M}{r^3f}\right)\right]=0~.
\end{equation}
With $M(r)$ being fixed at all times (from eq.(\ref{msm})), the vanishing of ${\dot f}$ at equilibrium
implies from eq.(\ref{eqom}) that 
\begin{equation}
G_e(r)=1-\frac{M(r)}{R_e(r)} = 1-\frac{M(r)}{rf_e(r)}~,
\label{geq}
\end{equation}
where the subscript $e$ denotes equilibrium values, and 
we have written $R_e(r) = r f_e(r)$. That this satisfies eq.(\ref{ddot}) is then readily checked. A few words about the
equilibrium scenario is in order. 

First, note that at equilibrium, all time derivatives of $f$ vanish. This is seen by differentiating the 
relation ${\dot f}^2 = -V(r,f)$. With ${\dot f} = {\ddot f} = 0$, this gives $f^{(3)} = -\frac{1}{2}V_{,f,f}{\dot f}=0$ and thus by induction,
it follows that $f^{(n)}=0$, where the subscript $(n)$ denotes the $n$th derivative of $f$ with respect to the co-moving time. 
From eq.(\ref{scaling}),
it is then clear that all time derivatives of $R(r,t)$ are zero at equilibrium, as expected. Now, since at equilibrium
$V=V_f=0$, close to equilibrium we can write $V(r,f) = h(r)\left(f-f_e\right)^2$, i.e 
${\dot f}^2 = -h(r)\left(f-f_e\right)^2$, with $h(r)$ being a negative definite function of $r$. Then it follows that in terms of
the positive definite function $Y(r) = \sqrt{-h(r)}$, we have (since ${\dot f}$ is negative),
\begin{equation}
{\dot f} = -Y(r)\left(f-f_e\right)~,
\label{ec1}
\end{equation}
where we have made use of the fact that at a given value of $r$, the equilibrium value of $f$ should be less than the 
neighbouring values, since $R(r,t) = rf(r,t)$ is a decreasing function of time. Integrating eq.(\ref{ec1}),
we then obtain \cite{Joshi:2011zm}
\begin{equation}
\left(f - f_e\right) = e^{-Y(r)\left(t_e - t_i\right)}~,
\label{evolution}
\end{equation}
where $t_e$ is the equilibrium value of the co-moving time and $t_i$ is an initial reference time. 
It is then seen that equilibrium is reached in a very large co-moving time. The picture that emerges is that the system
climbs slowly up the potential ``hill'' and reaches an equilibrium value after a very long time, so in this sense we 
expect the solution to be dynamically long lived, i.e a meta-stable state as we have pointed out in the
introduction. 

Since an equilibrium configuration is reached at a very large co-moving timescale, the final configuration 
is strictly not static at finite co-moving times. In view of the comments in the last paragraph, it is not unreasonable
to assume that towards the end stage of the evolution of our collapsing metric, the evolution is sufficiently slow to warrant
a static approximation. Indeed, since the function $f - f_e$ falls off exponentially with the co-moving time as in 
eq.(\ref{evolution}), one can envisage a sufficiently large neighborhood near the (large) equilibrium co-moving
time in which such an approximation should hold. It is in this sense that we will talk about a static configuration
in the discussion to follow.

We now comment on the energy conditions. 
In general, as we have mentioned, in order to avoid a shell-crossing singularity, we require $R' > 0$, 
so that the density $\rho$ of eq.(\ref{eineqns}) is positive definite. Now, from the definition
of the Misner-Sharp mass of eq.(\ref{ms}), we obtain
\begin{equation}
{\dot G} = \frac{M{\dot R}}{R^2} - 2{\dot \nu}e^{-2\nu}{\dot R}^2 + 2e^{-2\nu}{\dot R}{\ddot R}~.
\label{dotG}
\end{equation}
Close to equilibrium, since ${\dot R}$ and ${\ddot R}$ go to zero, and assuming that ${\dot \nu}$ is regular, we
can neglect the last two terms on the right hand side of eq.(\ref{dotG}) compared to the first term, 
to see that ${\dot G} < 0$. This then
implies from eq.(\ref{vanond}) that $\nu'>0$. Then, with $R'>0$, the expression of $P_{\perp}$ in eq.(\ref{eineqns}) is positive
near equilibrium. With $\rho$ and $P_{\perp}$ positive, we expect the strong and weak energy conditions 
to be valid close to equilibrium. 

It is to be noted that energy conditions for collapsing solutions with vanishing radial pressure at generic
co-moving times during the collapse has been
considered in details in e.g. \cite{Magli1}, \cite{Magli2}, \cite{JM2}. The situation here is however qualitatively
different, as the collapse halts at its end stage. Our system exemplifies collapse with non-vanishing 
acceleration $a_{\mu}$, expansion $\Theta$ and shear $\sigma$ 
with respect to a co-moving observer, and these are given by \cite{Stephani}
\begin{equation}
a_{\mu} = \nu'\delta^{r}_{\mu}~,~~
\Theta = \frac{e^{-\nu}}{2}\left(4\frac{{\dot R}}{R} + 2\frac{{\dot R'}}{R'} - \frac{{\dot G}}{G}\right)~,~~
\sigma^2 = \frac{e^{-2\nu}}{\left(4GRR'\right)^2}\left[\left(R{\dot G} + 2G{\dot R}  \right)R' - 2GR{\dot R'}\right]^2~.
\end{equation}
Here, we have defined the four-velocity of the co-moving observer $u^{\mu} = e^{-\nu}\delta^0_{\mu}$ whence
the acceleration is $a^{\mu} = u^{\mu}_{;\nu}u^{\mu}$. $\Theta$ and $\sigma$ are computed from the standard
textbook expressions for the expansion and the shear, which are not reproduced here. As is known in the
literature, the acceleration arises due to the non-zero tangential stresses (see eq.(\ref{eineqns})). Importantly,
remembering that $R = rf(r,t)$ (with $f_e \neq 0$ for $r \neq 0$), we see that 
the expansion and the shear go to zero at the end stage of collapse, when all the time derivatives become zero, 
everywhere except close to the singularity, which is expected. Of course,
if we assume that there is no shell-crossing singularity (so that $R'>0$) and also that $\nu'>0$, then all the standard
energy conditions will be satisfied during collapse along with the conditions on $\Theta$ and $\sigma$, 
but these assumptions might be difficult to justify throughout
the collapse process. 

\subsection{The final static configuration}

We are now ready to compute the static configuration discussed in the last subsection.
The fact that the radial pressure vanishes implies, from the metric $ds_+^2$ of eq.(\ref{both}), that at equilibrium, 
\begin{equation}\label{nuprime}
\frac{1}{R_e(r)^2}\left(G_e(r) -1 + \frac{2G_e(r)R_e(r)\nu_e'(r)}{R_e'(r)}\right)=0~,
~~{\rm i.e}~~\nu_e'(r) = \frac{\left(1 - G_e(r)\right)R_e'(r)}{2G_e(r)R_e(r)}~.
\end{equation}
Hence, once we specify the form of $R_e(r)$ (equivalently, $f_e(r)$), the above equation in turn can be used to
solve for $\nu'(r)$, and the equilibrium metric can be determined from eqs.(\ref{geq} and (\ref{nuprime}).

In what follows, following the discussion after eq.(\ref{densityeq}), we assume a power law behavior, 
\begin{eqnarray}
f_e(r) = b r^\alpha~,~{\rm i.e}~R_e(r) = br^{\alpha + 1}~,
\label{afer}
\end{eqnarray}
where $b$ carries dimensions of length and $\alpha \ge 0$ is a
dimensionless non-negative constant.\footnote{Negative value of $\alpha$ ($-1<\alpha<0$) 
are ruled out although these might represent regular solutions, since we find later
from eq.(\ref{presseq}) that for these, the central density is zero and increases in a radial direction. Such
values are therefore unphysical.} The value of $b$ cannot be arbitrarily set, and needs to be
constrained by a physical argument. This is due to the following reason. 
The coordinate intervals used to describe the
expanding space-time for the metric $ds_-^2$ of \eqn{both} are given by $0\le r \le 1$, $0 \le \theta \le
\pi$ and $0\le \phi < 2\pi$. The Friedman equation governing its
dynamics is given by,
\begin{eqnarray} 
\frac{H^2}{H_0^2}=\Omega_{m0}\left(\frac{a_0}{a}\right)^3 + (1-\Omega_{m0})\left(\frac{a_0}{a}\right)^2 
\label{fried}
\end{eqnarray}
where $H=\dot{a}/a$ is the Hubble parameter for the over dense
sub-universe and $H_0,\,a_0$ are the values of $H$ and $a$ just after
the over dense region detaches from the background 
expansion. Here $\Omega_{m0}=\rho_0/\rho_{c0}$ where $\rho_{c0} =
3H_0^2$ and $\rho_0$ is the matter density when $a=a_0$. 
The solution of \eqn{fried} can be given in
a parametric form \cite{Weinberg} :
\begin{eqnarray}
 a = {a_0 \Omega_{m0}\over 2 (\Omega_{m0}-1)} (1- \cos\theta),~~~
 t = {\Omega_{m0}\over 2 H_0 (\Omega_{m0}-1)^{3/2}} (\theta-\sin\theta)\, .
\label{at}
\end{eqnarray}
As the sub universe is over dense, we assume $\Omega_{m0}>1$.  In this
simple picture of spherical collapse, the over dense region expands
initially and the maximum value of the scale factor is attained
when
\begin{equation}\label{amax}
a_{\rm max}={a_0 \Omega_{m0}\over (\Omega_{m0}-1)}\,,\,\,\,\, t_{\rm max}={\pi 
\Omega_{m0}\over 2 H_0 (\Omega_{m0}-1)^{3/2}}\,.
\end{equation}
During the expansion phase of the dust, the scale factor $a$ is expressed in
terms of $a_0$ as shown in \eqn{at}. The dependence of $a_0$ is carried over to the expression of
$a_{\rm max}$. Initially when contraction starts, the physical scale factor is $R(r, t_{\rm max}) = r a_{\rm max}$ whose
value depends on the choice of $a_0$ and consequently the final equilibrium value of $f_e(r)$ is also
dependent on this choice. To eradicate this arbitrariness one
has to assign some value to $a_0$. If we demand that $a_{\rm max} = 1$, this
specifies $a_0 = (\Omega_{m0} - 1)/\Omega_{m0}$. Now, once we have used the scaling freedom to
fix $a_{\rm max}$, we do not have any further freedom left to choose the constant $b$ in \eqn{afer}. 
For the moment, we will retain the quantity $a_{\rm max}$ to make the dimensions apparent and later scale
this quantity to unity. 

Note that on physical grounds, a general constraint on $R(r,t)$ for a collapsing scenario 
should be $R_{\rm max} > R_e$. This implies that for all values of $r$ (upon scaling $a_{\rm max}$ to unity),
\begin{equation}
\frac{R_{\rm max}}{R_e} = \frac{a_{\rm max}}{br^{\alpha}} > 1 \Longrightarrow \left(\frac{b}{a_{\rm max}}\right) < 1~,
\label{ns1}
\end{equation}
where the second inequality follows from the fact that the radial coordinate $r$ can be taken arbitrarily close to unity.

Next, we note that during the start of the collapse, no apparent horizons are formed. 
For a spherically symmetric space-time $ds_+^2$ defined in \eqn{both}, the equation of the apparent
horizon takes the form \cite{Joshibook}
\begin{equation}
g^{\mu\nu}\partial_{\mu} R(r,t) \partial_{\nu} R(r,t) = 0 \Longrightarrow \frac{M(r,t)}{R(r,t)} = 1
\label{aphor}
\end{equation}
If $M/R < 1$, the formation of apparent horizons is ruled out, whereas it always forms for $M/R \geq 1$. 
In our case, the condition that an apparent horizon does not form at the beginning of the collapse translates to
\begin{equation}
\frac{M(r,t_{\rm max})}{R_{\rm max}} = \frac{r^3 a_{\rm max}}{r a_{\rm max}} = r^2 < 1~,
\end{equation}
which is always satisfied in the range of $r$ that we are considering during the start of collapse. In the collapsing stage,
when $R(r,t)$ decreases, we cannot make a comprehensive statement regarding $M/R$, in the absence
of an analytic solution valid throughout the collapse process. However, it is interesting to consider the 
issue of trapped surfaces at the end stage of collapse, i.e at equilibrium. From \eqn{aphor}, we obtain
the location of the apparent horizon 
\begin{equation}
\frac{M(r)}{R_e} = \frac{a_{\rm max} r^3}{br^{\alpha+1}}=1 \Longrightarrow r^{2-\alpha} = \left(\frac{b}{a_{\rm max}}\right) ~.
\label{aphor1}
\end{equation}
From the above equation we see that for $\alpha = 2$, the apparent horizon
condition implies that $b/a_{\rm max}=1$ which is ruled out. Also note that $\alpha > 2$ implies that there are no apparent horizons 
for any value of $r$. However, this last case can also be ruled out on physical grounds, as we will see shortly. 

If we input the forms of $G_e(r)$ from \eqn{geq} and $R_e(r)$ from \eqn{afer}, then the second equation
in eq.(\ref{nuprime}) can be immediately
integrated to obtain the $g_{00}$ component of the equilibrium metric. The $g_{11}$ component is also easy 
to obtain, and we finally find the equilibrium metric (we write this in the $r$ coordinate, but this can always be 
converted to the physical $R_e$ coordinate by the transformation $R_e = br^{\alpha + 1}$) given by :
\begin{equation}
ds_e^2 = -{\mathcal A}\left(\frac{b}{a_{\rm max}}-r^{2-\alpha }\right)^{\frac{\alpha +1}{\alpha -2}}dt^2 + 
\frac{(\alpha +1)^2 b^3 r^{2 \alpha }}{b- a_{\rm max} r^{2-\alpha }}dr^2
+ b^2 r^{2\left(\alpha +1\right)}d\Omega^2~.
\label{meteq}
\end{equation}
Here, ${\mathcal A}$ is an integration constant (whose value we will momentarily fix). The energy momentum tensor is evaluated 
to be ($P_r=0$ by construction)
\begin{equation}
\rho_e = \frac{3a_{\rm max}r^{-3\alpha}}{b^3\left(1+\alpha\right)},~~
P_{\perp,e} = \frac{3a_{\rm max}r^{2-4\alpha}}{4b^3\left(\frac{b}{a_{\rm max}} - r^{2-\alpha}\right)\left(1+\alpha\right)}~.
\label{presseq}
\end{equation}
On physical grounds, we see that the metric of \eqn{meteq} is valid for $r < (b/a_{\rm max})^{1/(2-\alpha)}$. Since we have already
seen that $(b/a_{\rm max}) < 1$, this implies that $\alpha < 2$,\footnote{If $\alpha \geq 2$, then, for example, $g_{11}$ 
becomes negative, since $b/a_{\rm max}<1$ and $0\leq r<1$. 
The tangential pressures also become negative in this case, as can be seen from \eqn{presseq}.}
so that we obtain a physical metric in the range of $r$ from $r = 0$ to $r = (b/a_{\rm max})^{1/(2-\alpha)}$, i.e
up to the location of the apparent horizon. We can therefore use the following parameter set for the discussion to follow, 
\begin{equation}
0\leq\alpha < 2,~~0<\frac{b}{a_{\rm max}}<1~.
\label{con1}
\end{equation}

At this point, in order not to clutter notations, we will use the scaling $a_{\rm max}=1$ discussed earlier, and
use numerical bounds on $b$, as a convenience. 
The solution of eq.(\ref{meteq}), along with eq.(\ref{con1}) thus describes a two-parameter family of static solutions. 
As we will see later, one can further restrict the parameter $\alpha$ from physical arguments. 
With the range of parameters specified by eq.(\ref{con1}), the tangential pressures of \eqn{presseq} are 
positive definite. 
The weak energy conditions are automatically satisfied in this range, given 
the positivity of the density and the pressures here, 
as is the strong energy condition $\rho + 2P_{\theta} > 0$. At this point, we remember that as mentioned in
the beginning of this section (see discussion after eq.(\ref{msm})), we have not supplied an
equation of state for the collapsing matter, and this is determined from the end state of the collapse. Indeed, 
as can be seen from eq.(\ref{presseq}), our solution of eq.(\ref{meteq}) satisfies a complicated ``equation of state''
for the tangential pressures, given by 
\begin{equation}
P_{\perp,e} =\frac{\rho_e}{4\left(b\left(\frac{\rho_e}{k_1}\right)^{\frac{2-\alpha}{3\alpha}}-1\right)}~,~{\rm with}~~
k_1 = \frac{3}{b^3\left(1+\alpha\right)}~.
\label{eosinitial}
\end{equation}
It is therefore difficult to envisage the matter as a simple barotropic fluid akin to some form of dark matter. 
It is interesting to investigate a two-fluid description of the same. We will come back to this in a while. 

Now, the Ricci scalar ($Ric$) is computed to be
\begin{equation}
Ric = \frac{3 r^{-4 \alpha } \left(2 b r^{\alpha }-3 r^2\right)}{2 (\alpha +1) b^3 \left(b-r^{2-\alpha}\right)}~.
\label{Ric}
\end{equation}
As can be seen, apart from the singularity at $r=0$ (for $\alpha >0$), 
there is a singularity at the location of the apparent horizon
given by \eqn{aphor1}.\footnote{The Kretschmann scalar is also singular at the origin and at the location of the apparent
horizon.} However, this is not of concern, as we 
have already assumed that the metric of \eqn{meteq} is valid for $r < b^{1/(2-\alpha)}$
and that it matches smoothly to a Schwarzschild 
solution at some matching radius 
\begin{eqnarray}\label{con2}
0<r_m < b^{1/(2-\alpha)}\, .
\end{eqnarray} This type of smooth 
matching without a matter shell at the boundary is always possible 
in principle, as the radial component of the pressure is zero. 
Doing this determines the value of ${\mathcal A}$ in \eqn{meteq}, 
and converting to the physical $R_e$ ($=br^{\alpha + 1}$) coordinate 
with $R_m$ being a matching radius, we present our final form of the equilibrium metric :
\begin{equation}
ds_e^2 = - \left( 1-\frac{1}{b}\left(\frac{R_m}{b}\right)^{\frac{2-\alpha}{1+\alpha}}\right)
\left(\frac{1 - \frac{1}{b}\left(\frac{R_m}{b}\right)^{\frac{2-\alpha}{1+\alpha}}}
{1 -\frac{1}{b} \left(\frac{R_e}{b}\right)^{\frac{2-\alpha}{1+\alpha}}}\right)^{\frac{1+\alpha}{2-\alpha}}dt^2 
+ \frac{dR_e^2}{1-\frac{1}{b}\left(\frac{R_e}{b}\right)^{\frac{2-\alpha}{1+\alpha}}} + R_e^2d\Omega^2~,
\label{dsfinal}
\end{equation}
where the metric of \eqn{dsfinal} is valid up to 
\begin{eqnarray}\label{con3a}
 R_e < b^{3/(2-\alpha)}\, .
\end{eqnarray} 
This metric is matched to a Schwarzschild metric at $R_e = R_m$, with mass 
\begin{equation}
M = \frac{1}{2}\left(\frac{R_m}{b}\right)^{\frac{3}{1+\alpha}} = \left(1+\alpha\right) \frac{4}{3}\pi R_m^3 \rho_e~,
\label{mass}
\end{equation}
where the last identity follows from the expression of the equilibrium value of the energy density 
$\rho_e$ (at $R_e = R_m$) obtained from the metric of \eqn{dsfinal}. 

A few words about this metric is in order. Setting $\alpha = 0$, we obtain
\begin{equation}
ds_e^2 = -\frac{\left(1 - \frac{R_m^2}{b^3}\right)^{\frac{3}{2}}}{\sqrt{1 - \frac{R_e^2}{b^3}}}dt^2 + \frac{dR_e^2}{1-\frac{R_e^2}{b^3}} + R_e^2d\Omega^2~,
\end{equation}
and one can recognize this to be the interior Schwarzschild solution of Florides \cite{Florides}.
In this case, there is no
curvature singularity at $R_e = r = 0$ as can be seen from \eqn{Ric}. The only such singularity occurs at 
$R_e = b^{3/2}$. The metric is therefore regular in the interval $0\leq R_e<b^{3/2}$. The energy density $\rho_e$ is constant
in this case, as can be gleaned from \eqn{presseq}. Then, the mass $M$ of eq.(\ref{mass}) is the total mass contained
in a sphere of radius $R_m$, as expected. 

For values of $\alpha > 0$, there is a singularity at the origin, as well as one at $r = b^{1/(2-\alpha)}$ or equivalently $R_e = b^{3/(2-\alpha)}$. 
As already mentioned, we will not be concerned with the latter singularity, and we assume that the metric of \eqn{dsfinal} will be matched to an 
external Schwarzschild solution before we reach this value of $R_e$. 
That the singularity at the origin is naked can be seen from the following. We take the metric of \eqn{dsfinal} and compute 
the time $\tau$ taken for null geodesics to reach a value of $R_e$ starting from the origin. We find that this time equals
\begin{equation}
\tau = b^{\frac{7+\alpha}{4-2\alpha}}\, _2F_1\left(\frac{1-2 \alpha }{4-2 \alpha },\frac{\alpha +1}{2-\alpha};\frac{3}{2-\alpha };
\left(\frac{R_e^{2-\alpha}}{b^3}\right)^{\frac{1}{\alpha + 1}} \right)~,
\label{time1}
\end{equation}
where $_2F_1$ is the Gauss hypergeometric function. 
That the time to reach the boundary at $R_e = b^{3/(2-\alpha)}$ is finite is then evident. 

Finally, a word on the dominant energy condition, given in general by $P_{\theta,e}/\rho_e < 1$, required to rule out the possibility of 
superluminal propagation of sound. It can be seen from \eqn{presseq} that in order to satisfy this condition, we require that 
\begin{equation}
\frac{P_{\perp,e}}{\rho_e} < 1 \Longrightarrow \left(R_e^{\alpha - 2}b^3\right)^{\frac{1}{\alpha + 1}} > \frac{5}{4}~.
\end{equation}
This would in general imply that 
\begin{equation}
R_e < \left(\frac{4}{5}\right)^{\frac{1+\alpha}{2-\alpha}} b^{\frac{3}{2-\alpha}}~.
\label{DEC}
\end{equation}
This is a somewhat stronger condition than $R_e < b^{3/(2-\alpha)}$, that we saw earlier for the validity of the metric of \eqn{dsfinal}. 
The validity of the dominant energy condition can be ensured by choosing the matching radius (with the external Schwarzschild
metric) at a value $R_m$ less than or equal to the right hand side of the inequality in \eqn{DEC}. Note that 
at the matching radius dictated by eq.(\ref{DEC}), the density and the tangential pressures are equal, as follows
readily from eq.(\ref{presseq}). We will remember this fact which will be useful later. 

Now, as we had mentioned earlier, we can further restrict the parameter range for $\alpha$ presented in eq.(\ref{con1}). This can 
be done as follows. On physical grounds, it is reasonable to demand that the gradient of the tangential pressure 
of eq.(\ref{presseq}) should be negative definite, i.e the pressure always decreases as one moves away from
the center, as is the case for the JMN solution of \cite{Joshi:2011zm}. This dictates that $\alpha > 1/2$, and further 
can be shown to require the condition
\begin{equation}
R_e < \left(\frac{4\alpha - 2}{3\alpha}\right)^{\frac{1+\alpha}{2-\alpha}}b^{\frac{3}{2-\alpha}}~.
\label{rFinal}
\end{equation}
The maximum radius obtained in eq.(\ref{rFinal}) should be less than the one obtained in eq.(\ref{DEC})
to prevent superluminal propagation of sound. 
In order to enforce this condition, we obtain a restricted range of
parameters, 
\begin{equation}
\frac{1}{2}<\alpha < \frac{5}{4},~~0<b<1~.
\label{con3}
\end{equation}
This is our final constraint on the parameters $\alpha$ and $b$ that specify the solution of eq.(\ref{dsfinal}). The matching
radius $R_m$ of eq.(\ref{dsfinal}) has to be chosen such that $R_m$ is less than $R_e$ given in eq.(\ref{rFinal}). 
From our previous discussion, it follows that the non-central singularity (see eq.(\ref{con3a})) is always avoided
by our choice of parameters.

It is important to contrast our result of eq.(\ref{dsfinal}) 
with that of \cite{Joshi:2011zm} where a different naked singularity, the JMN singularity
was obtained. That solution was done for the specific choice $b=1$ and $\alpha = 2$. Both these values are ruled out in our
case, as seen from eq.(\ref{rFinal}), and the analysis there needs to be done separately, and indeed leads to the JMN solution. 
We will recall a few salient features of our analysis. First, the form of the Misner-Sharp
mass function is fixed in our analysis by the matching conditions (see eq.(\ref{msm})) and does not need to be supplied
by hand. Second, since we start with an
initial FLRW metric, the parameter $b$ is constrained by eq.(\ref{ns1}) and due to this, the parameter $\alpha$ is constrained
to be less than $2$ (see discussion after eq.(\ref{presseq})). For the JMN solution, one leaves $a_{\rm max}$ free, 
and fixes $b=1$ and $\alpha = 2$. With these values, $a_{\rm max}$ determines a one-parameter family of
solutions. Our solution is therefore qualitatively different from the
JMN space-time and represent a new class of naked singularities that can arise out of gravitational collapse. 
As we will show in the next section, there are therefore important differences in gravitational lensing and
accretion disk properties of our model, compared to the JMN solution.

Before ending this subsection, we point out that we have not studied the stability of the solution
described in eq.(\ref{dsfinal}). A quasi-normal mode analysis is difficult to envisage here, in the absence of 
a horizon. In \cite{Suneeta}, the stability of the JNW naked singularity under a perturbation of the sourcing scalar
field was studied. It will be interesting to perform such an analysis in the context of the fluid sourcing our space-time.
It is also to be noted that although we have scaled $a_{\rm max}$ of eq.(\ref{msm}) to unity in our analysis, any other
value of this scale factor would not affect the metric of eq.(\ref{dsfinal}), and would simply rescale the parameters
$\alpha$ and $b$.

\subsection{The fluid sourcing the final configuration}

We now discuss the important issue of the nature of the fluid sourcing our space-time of eq.(\ref{dsfinal}). 
As pointed out before, there is no simple equation of state corresponding to the matter content, as is evident
from eq.(\ref{eosinitial}). However, we can envisage a scenario where the energy-momentum tensor 
is a combination of two non-interacting simple fluids that are barotropic in nature. In cosmology, these types
of models have a long history (see, e.g. \cite{Cohen}, \cite{Nowotny}, \cite{Coley}) where two fluids are 
assumed a-priori. Here, we will closely follow
the notations and conventions of \cite{Letelier}, \cite{Bayin} (see also \cite{BDS}), where it was shown that 
the energy momentum tensor of two non-interacting fluids can be written as one of a composite fluid.  
We remind the reader that we will work in the static approximation mentioned before and assume that
the energy and momenta of the two fluids are conserved separately.

This energy-momentum tensor (assumed to comprise of two noninteracting perfect fluids), is given by
\begin{eqnarray}
T_{\mu \nu} = (p_1 + \rho_1)u_\mu u_\nu + p_1 g_{\mu \nu} + (p_2 +
\rho_2)v_{\mu} v_\nu + p_2 g_{\mu \nu}\,,
\label{2femt}
\end{eqnarray}
where 
$$u^\mu u_\mu = v^\mu v_\mu = -1\,.$$ Now, one conventionally changes the basis in the 4-velocity space \cite{Letelier}, 
so that $T_{\mu\nu}$ is cast to a standard form, 
\begin{eqnarray}
T_{\mu \nu} = (\rho + P_\perp)w_\mu w_\nu + P_\perp g_{\mu \nu} +
(P_r - P_\perp) y_\mu y_\nu\,,
\label{2frmt1}
\end{eqnarray}
where 
$w^\mu w_\mu =-1\,,\,\,\,y^\mu y_\mu=1\,,\,\,\,w^\mu y_\mu =0\,,$ Here, $w_\mu$ is the 4-velocity of the effective two-fluid
system and $y_\mu$ is a spacelike vector along the anisotropy direction. The energy densities and pressures of the
component fluids are denoted by $\rho_1, p_1$ and $\rho_2, p_2$ respectively. In terms of these, it can be shown
\cite{Letelier}, \cite{Bayin} that the components of the energy-momentum tensor of the composite fluid is 
\begin{eqnarray}
\rho &=& \frac{1}{2}(\rho_1 - p_1 + \rho_2 - p_2) + \frac{1}{2} \left[(p_1 + \rho_1 - p_2 - \rho_2)^2
+ 4(u_\mu v^\mu)^2(p_1 + \rho_1)(p_2 + \rho_2)\right]^{1/2}\,,\nonumber\\
P_r &=&-\frac{1}{2}(\rho_1 - p_1 + \rho_2 - p_2) + \frac{1}{2} \left[
(p_1 + \rho_1 - p_2 - \rho_2)^2 + 4(u_\mu v^\mu)^2(p_1 + \rho_1)(p_2 + \rho_2)\right]^{1/2}\nonumber\\
P_\perp &=& p_1 + p_2\,.
\label{prescomp}
\end{eqnarray}
For spherically symmetric anisotropic fluids, one conventionally chooses
\begin{eqnarray}
y^0=y^2=y^3=0~;~~~~~w^1=w^2=w^3=0
\end{eqnarray}
so that $y^1y_1= 1$ and $w^0w_0=-1$. Then, we have $T^0_0 = -\rho,~T^1_1 = P_r,~~T^2_2 = T^3_3 = P_{\perp}$.

Now, the Einstein equations give three equations in terms of five unknowns, i.e 
$\rho_1(r)$, $\rho_2(r)$, $p_1(r)$, $p_2(r)$, and $u_\mu v^\mu(r)$. If one assumes the simple barotropic 
equations of state for the two fluids
\begin{eqnarray}
p_1 = \gamma_1 \rho_1~,~~~p_2=\gamma_2 \rho_2~,
\label{eos2}
\end{eqnarray}
where $\gamma_1$ and $\gamma_2$ are constants, then we can solve these equations. 
In our case (with $P_r=0$), we can solve eq.(\ref{prescomp}) to obtain
\begin{equation}
K \equiv \left(u_\mu v^\mu\right)^2= \frac{\rho_1\rho_2\left(1+\gamma_1\gamma_2\right)-
\left(\gamma_1\rho_1^2 + \gamma_2\rho_2^2\right)}{\rho_1\rho_2\left(1+\gamma_1\right)\left(1+\gamma_2\right)}~,
\label{Keq}
\end{equation}
where we have to choose $u_\mu v^\mu$ with a negative sign, as $u$ and $v$ are assumed to be time-like vectors. 
The individual densities can be shown to be given by \cite{BDS}
\begin{equation}
\rho_1 = \frac{\gamma_2\left(P_{\perp}  + \rho \right)-P_{\perp}}{(\gamma_2 - \gamma_1)}~,~
\rho_2 = \frac{\gamma_1(\rho+P_{\perp}) - P_{\perp}}{(\gamma_1 -\gamma_2)}~.
\label{densities}
\end{equation}
Using eq.(\ref{presseq}) after replacing $r$ by $R_e$ via $R_e(r) = br^{\alpha + 1}$, it is then seen that these reduce to 
\begin{equation}
\rho_1 = \frac{12 b \gamma _2 \left(\frac{R_e}{b}\right)^{\frac{\alpha +3}{\alpha +1}}-3
\left(3 \gamma _2+1\right) \left(\frac{R_e}{b}\right)^{\frac{5}{\alpha
+1}}}{4 (\alpha +1) \left(\gamma _1-\gamma _2\right) R_e^3
\left(\left(\frac{r}{b}\right)^{\frac{2}{\alpha +1}}-b
\left(\frac{R_e}{b}\right)^{\frac{\alpha }{\alpha +1}}\right)}~,~~
\rho_2 = \frac{3 \left(3 \gamma _1+1\right) \left(\frac{R_e}{b}\right)^{\frac{5}{\alpha
+1}}-12 \gamma _1 r \left(\frac{R_e}{b}\right)^{\frac{2}{\alpha +1}}}{4
(\alpha +1) \left(\gamma _1-\gamma _2\right) R_e^3
\left(\left(\frac{R_e}{b}\right)^{\frac{2}{\alpha +1}}-b
\left(\frac{R_e}{b}\right)^{\frac{\alpha }{\alpha +1}}\right)}~.
\label{rho12eq}
\end{equation}
Physical solutions will correspond to values of $\gamma_1$, $\gamma_2$ (with chosen values of $\alpha$ and $b$),
for which $\rho_1$, $\rho_2$, and $K$ of eq.(\ref{Keq}) are positive. The general analysis using 
eq.(\ref{rho12eq}) can be cumbersome, and we will only present some specific simple examples. 
For example, if we take $\gamma_1 = 0$ so that one of the component fluids is dust, 
then eqs.(\ref{Keq}) and (\ref{densities}) reduces to 
\begin{equation}
\rho_1 = \rho + P_{\perp}\left(1-\frac{1}{\gamma_2}\right)~,~~\rho_2 = \frac{P_{\perp}}{\gamma_2}~,~~
K = \frac{\rho_2\left(\rho_1 - \gamma_2\rho_2\right)}{\rho_1\rho_2\left(1+\gamma_2\right)}~.
\label{special}
\end{equation}
Now suppose we set $\gamma_2=1$, then the other fluid follows a simple barotropic 
equation of state $p_2=\rho_2$, which is common in the fluid description of a free scalar field, also called
stiff matter. Then, from eq.(\ref{special}), it is seen that $\rho_1=\rho$ and 
$\rho_2 = P_{\perp}$. The density of dust is then the density of the composite fluid, and the pressure anisotropy 
acts as the density of the non-dust stiff matter. It remains to check if the right hand side of eq.(\ref{Keq}), whose 
numerator is given (with our choice of $\gamma_1=0$ and $\gamma_2=1$) by $P_{\perp}\left(\rho -P_{\perp}\right)$ 
is positive definite, since the denominator of that expression is positive definite. 
\begin{figure}[h!]
\begin{minipage}[b]{0.5\linewidth}
\centering
\includegraphics[width=3.2in,height=2.5in]{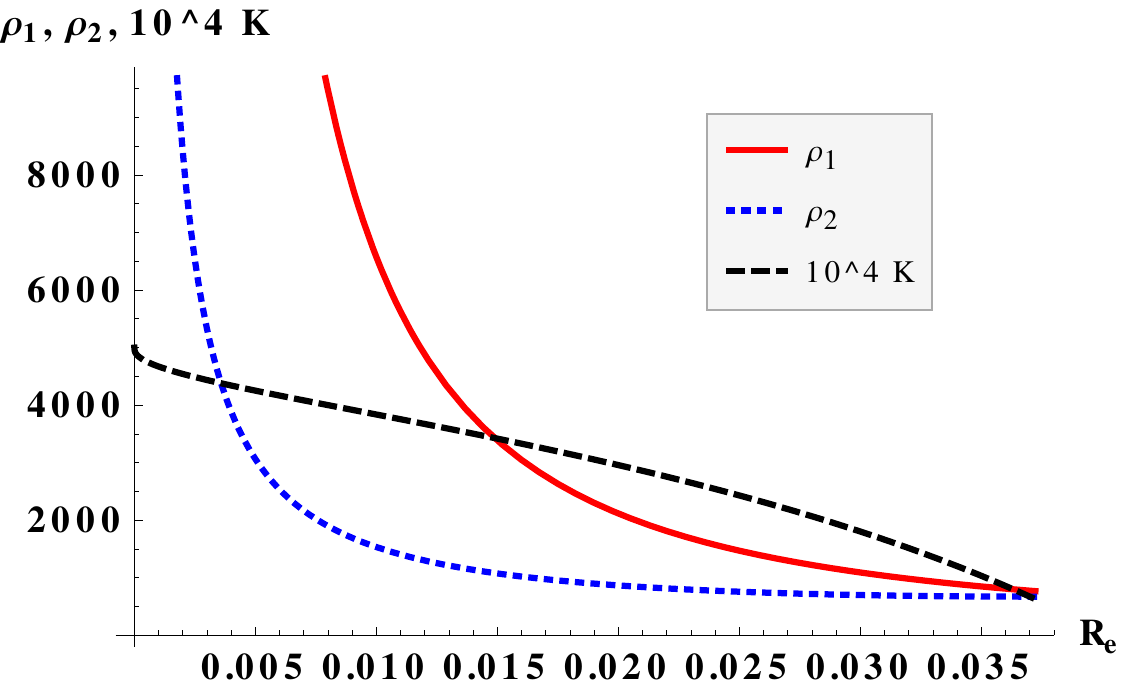}
\caption{$\rho_1$, $\rho_2$ and $10^4 K = 10^4 \left(u_\mu
  v^\mu\right)^2$ for $\alpha = 1.2$ and $b = 0.5$ for $\gamma_1=0$
  and $\gamma_2=1$.}
\label{twofluid1}
\end{minipage}
\hspace{0.4cm}
\begin{minipage}[b]{0.5\linewidth}
\includegraphics[width=3.2in,height=2.5in]{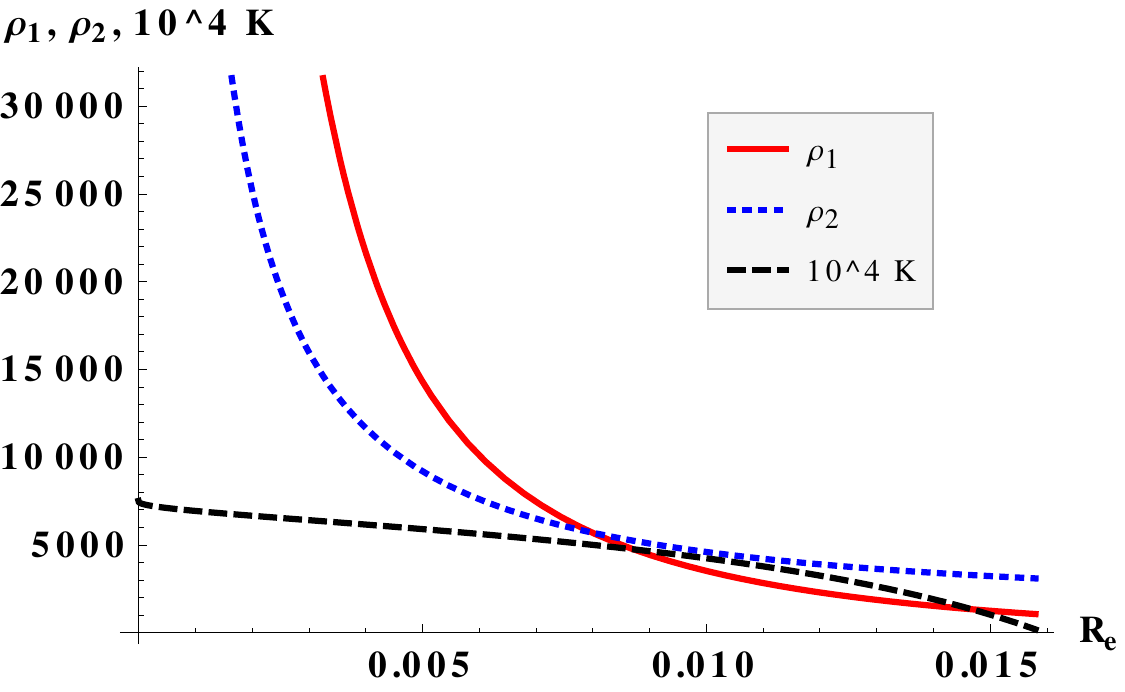}
\caption{$\rho_1$, $\rho_2$ and $10^4 K = 10^4 \left(u_\mu
  v^\mu\right)^2$ for $\alpha = 1.2$ and $b = 0.5$ for $\gamma_1=0$
  and $\gamma_2=1/3$.}
\label{twofluid2}
\end{minipage}
\end{figure}
We will choose the boundary of the space-time defined in eq.(\ref{rFinal}). We also set, 
for illustration, $\alpha = 1.2$, $b = 0.5$, which are consistent with the range given in eq.(\ref{con3}) 
and these values will be used subsequently as well. 
In fig.(\ref{twofluid1}), we have shown the quantities $\rho_1$,
$\rho_2$ and $10^4 K = 10^4 \left(u_\mu v^\mu\right)^2$ for these values (here, $K$ has been multiplied by a factor
of $10^4$ to offer better visibility). All quantities are finite at the boundary, and
$\rho_1$, $\rho_2$ diverge at the origin, a fact that reflects the original singularity. Further, these
quantities decrease as one moves away from the origin, which is physically reasonable. 
Fig.(\ref{twofluid2}) shows the corresponding situation with the barotropic index $\gamma_2 = 1/3$ with $\gamma_1=0$.
In this case, one of the component fluids is dust and the other is pure radiation. 
Here, we need to fix the matching radius at a value lower than that dictated by eq.(\ref{rFinal}), which can be obtained by solving 
the equation $K=0$. Qualitatively we see the same picture as presented in fig.(\ref{twofluid1}). 

The overall picture that emerges here is as follows. As we have already mentioned, 
the original fluid of eq.(\ref{dsfinal}) does not have a simple equation of state. However, in a two-fluid description, 
we are able to model this as a dust part, along with a simple barotropic fluid. One can consider this latter
fluid to be stiff matter, but there are other possibilities as well, and one might choose a barotropic index less than unity, in which
case, the matching radius has to be adjusted appropriately so that $K$ remains positive throughout the range
of the radial coordinate. 
We also note that the analysis above is quite general. For example, any fluid model with $P_r=0$ can be represented 
as a combination of dust and a stiff fluid with unit barotropic index, as long as $\rho - P_{\perp} >0$ in the region of interest. 
In all such cases, the dust part carries the information of the energy density of the composite fluid and the energy
density of the stiff matter is the pressure anisotropy of the original model. We should mention here that
it would be interesting to compute and analyze the peculiar velocities of the two fluids studied here (and possibly in 
two-fluid models of generic naked 
singularity backgrounds), via the method 
described in \cite{Bayin}. We have not performed this analysis here and this will be reported elsewhere. 

Having established the physical nature of the fluid sourcing our naked singularity of
eq.(\ref{dsfinal}), it now remains to elaborate upon some relevant observational properties for our model. 

\section{Observational properties : gravitational lensing and accretion disk} \label{physical}

In this section, we will carry out an analysis of some relevant observational 
properties of our solution, under the static approximation discussed in the previous section. 
In particular, we will focus on strong gravitational lensing and accretion disk
properties in this background. 
To this end, we will start by examining circular 
trajectories of massive particles here. We assume here that the matter seeding the metric of 
eq.(\ref{dsfinal}) does not interact with such particles, but merely feel the geometry. To simplify notations and 
for ease of reading, we will henceforth label $R_e = q$, and remember that $q$ is the radial coordinate of our
equilibrium metric in what follows. 

\subsection{Geodesics and photon orbits}

Since our equilibrium metric is spherically symmetric, we will, without loss of generality, choose the polar 
angle $\theta = \pi/2$. We recall the textbook result that for a generic static, spherically symmetric metric with components
$g_{tt}(q)$, $g_{qq}(q)$ and $g_{\phi\phi}(q)$, massive particle trajectories are described by the geodesic equation
\begin{equation}
{\dot q}^2 + V\left(q\right) = 0 ~,~~V(q) = \frac{1}{g_{qq}(q)}\left[\frac{E^2}{g_{tt}(q)} + \frac{L^2}{g_{\phi\phi}(q)} + 1\right]~,
\label{veff}
\end{equation}
where $E$ and $L$ denote the energy and angular momentum of the particle per unit mass. 
For stable circular orbits, since $V(q) = V'(q) = 0$ (along with $V''(q) >0$), we have
\begin{equation}\label{el}
E = \left[\frac{g_{tt}^2g_{\phi\phi}'}{g_{\phi\phi}g_{tt}' - g_{tt}g_{\phi\phi}'}\right]^{\frac{1}{2}}~,~~
L = \left[-\frac{g_{\phi\phi}^2g_{tt}'}{g_{\phi\phi}g_{tt}'- g_{tt}g_{\phi\phi}'}\right]^{\frac{1}{2}}~,
\end{equation}
Also note that the angular speed of the particle is given as 
\begin{equation}\label{omnew}
\omega = \frac{d\phi}{dt} = \left[-\frac{g_{tt}'}{g_{\phi\phi}'}\right]^{\frac{1}{2}}~.
\end{equation}
It can be checked that for the metric of eq.(\ref{dsfinal}),  eq.(\ref{el}) can equivalently be written as 
\begin{equation}\label{le}
 E =\sqrt{-2 g_{tt}}\left(1-x\over 2-3x\right)^{1\over 2}~,~~ L = {q\sqrt x\over\sqrt{2-3x}}~,
\end{equation}
where $x ={1\over b}(\frac{q}{b})^{2-\alpha\over 1+\alpha}$. 
Note that near the origin, i.e when $q\to 0$, the angular momentum per unit mass goes to zero, but the energy
reaches a value $E_{q\to 0} = \sqrt{-g_{tt}}$. This affects the radiative efficiency of the model, as will be discussed
at the end of subsection (5.2).
\begin{figure}
\centering
 \includegraphics[width=4.00in]{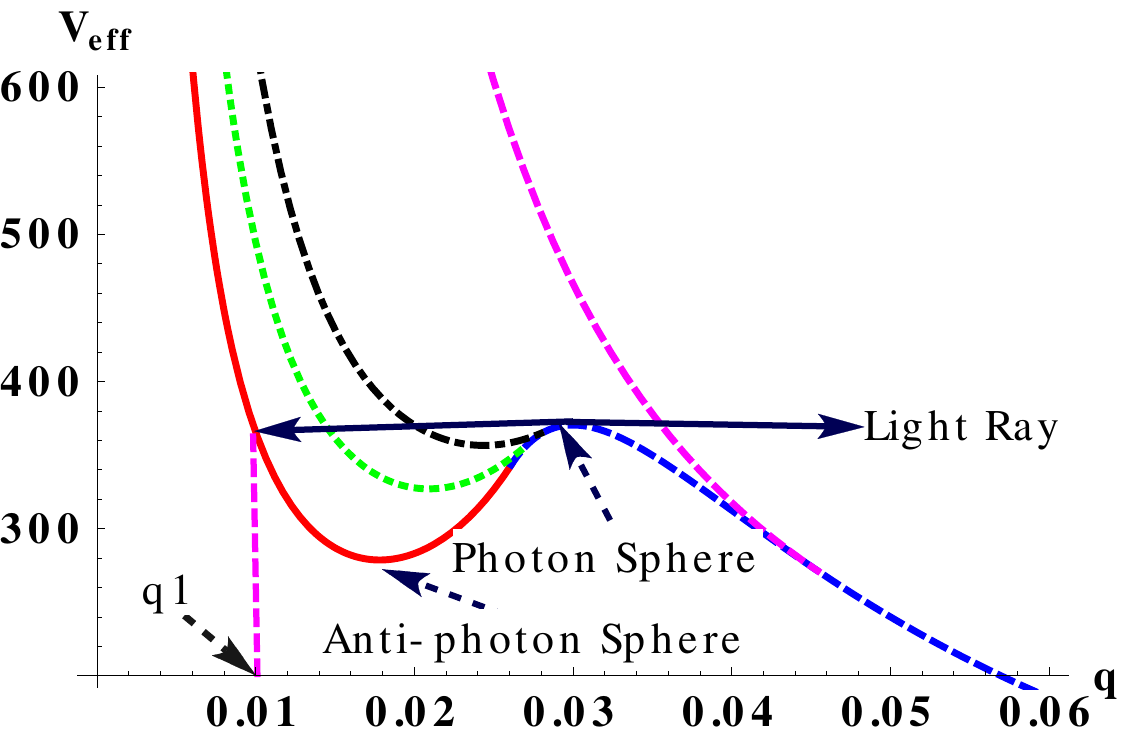}
 \caption{Photon effective potential with $L_{ph}=1$ 
 for $M=0.01$ and $\alpha=1.2$. The solid red, dotted green, dot dashed-black and dashed 
pink lines correspond to $b=0.46$, $0.48$, $0.50$ and $0.80$ respectively. The dashed blue line represents the external
Schwarzschild solution with $M=0.01$. The horizontal solid black line represents 
a light ray (see text). }
 \label{fig1}
\end{figure}

Importantly, from eq.(\ref{le}), we note that $x$ should be less than $2/3$ for a physical stable circular orbit. This means
that although the metric of eq.(\ref{dsfinal}) is valid up to the radial distance defined in eq.(\ref{DEC}), 
circular orbits exist only up to a radial distance for an ``outermost stable circular orbit'' (OSCO) defined by 
\begin{eqnarray}
q < \left(2\over 3\right)^{1+\alpha\over 2-\alpha} b^{3\over 2-\alpha} = q_{\rm max}\, .
\label{qmax}
\end{eqnarray}
We also note that with the metric of eq.(\ref{dsfinal}), we get from eq.(\ref{veff}) and eq.(\ref{el})
\begin{equation}
V''(q) = \frac{2}{3q^3(1+\alpha)}\left(\frac{q}{b}\right)^{\frac{3}{1+\alpha}}\left[\frac{3x(1+\alpha) - (4+\alpha)}{x-\frac{2}{3}}\right]~.
\label{maxmin}
\end{equation} 
Hence, whenever $x<2/3$, we have $V''(q) > 0$, i.e there exists stable circular
orbits for massive particles at all radii up to the OSCO at radius $q_{max}$. 

Now we discuss photon orbits. It is useful to write the geodesic equation for the 
photon (after setting $\theta = \pi/2$) analogous to eq.(\ref{veff}) in terms of an effective potential $V_{eff}$, 
\begin{equation}
-g_{tt} g_{qq}{\dot q}^2 + V_{eff}= E_{ph}^2~,~V_{eff}=-L_{ph}^2\frac{g_{tt}}{g_{\phi\phi}}~.
\end{equation}
Now, for the metric of eq.(\ref{dsfinal}), circular photon orbits can occur \cite{V1},\cite{V2} 
at the solutions of 
\begin{equation}
2g_{tt} -qg_{tt}' = 0~,{\rm i.~e}~~q = q_{\rm max}~.
\label{ps}
\end{equation}
Hence circular photon geodesics are possible only at the value of $q_{\rm max}$ defined in eq.(\ref{qmax}). 
However, it can be checked that these are {\em stable} photon orbits, dubbed anti-photon spheres 
(or stable light rings) in the literature, where $V_{eff}''(r) > 0$. This is a crucial difference of our model
as compared to the one considered in \cite{Joshi:2011zm} in which there are no such stable photon orbits. 
Now, if the interior solution of eq.(\ref{dsfinal}) is matched with an external
Schwarzschild solution at $2M<R_m<3M$, then these stable photon sphere, along with the 
Schwarzschild photon sphere (at $q=3M$) play a critical role in the analysis and completely alter the
gravitational lensing pattern of our solution, compared to the ones obtained using the JMN solution. 

To be concrete, we will make choices of variables according to the previous discussion, and 
for a given constant energy $E_{ph}$ and constant angular momentum $L_{ph}$ of the photon, we first study the 
effective potential. In fig.(\ref{fig1}), we show the quantity $V_{eff}/L_{ph}^2$ (with $L_{ph}=1$)
with $M=0.01, \alpha=1.2$ for $b=0.46$ (solid red), $0.48$ (dotted green), $0.50$ (dot-dashed black) and $0.80$
(dashed pink). The dashed blue line shows the effective potential for an external Schwarzschild metric with $M=0.01$. 
In all these cases, the matching radius computed from eq.(\ref{mass}) is less than the physical range of the radial
coordinate given in eq.(\ref{con3}). 
For the first three cases, there is an anti-photon sphere, as the matching radius $R_m$ of
eq.(\ref{mass}) lies between $2M$ and $3M$. In the last case, $R_m > 3M$, resulting in no extrema of the
effective potential. In the figure we also schematically show the possible trajectory of a light ray close to the 
photon sphere which will be reflected back
when it hits the potential barrier. We also show by a vertical magenta line
a typical point $q_1$ (for $b=0.46$), where the height of the effective potential is the same as that of the photon sphere.
This will be useful to us in what follows. 
\begin{figure}[h!]
\begin{minipage}[b]{0.5\linewidth}
\centering
\includegraphics[width=3.2in,height=3.0in]{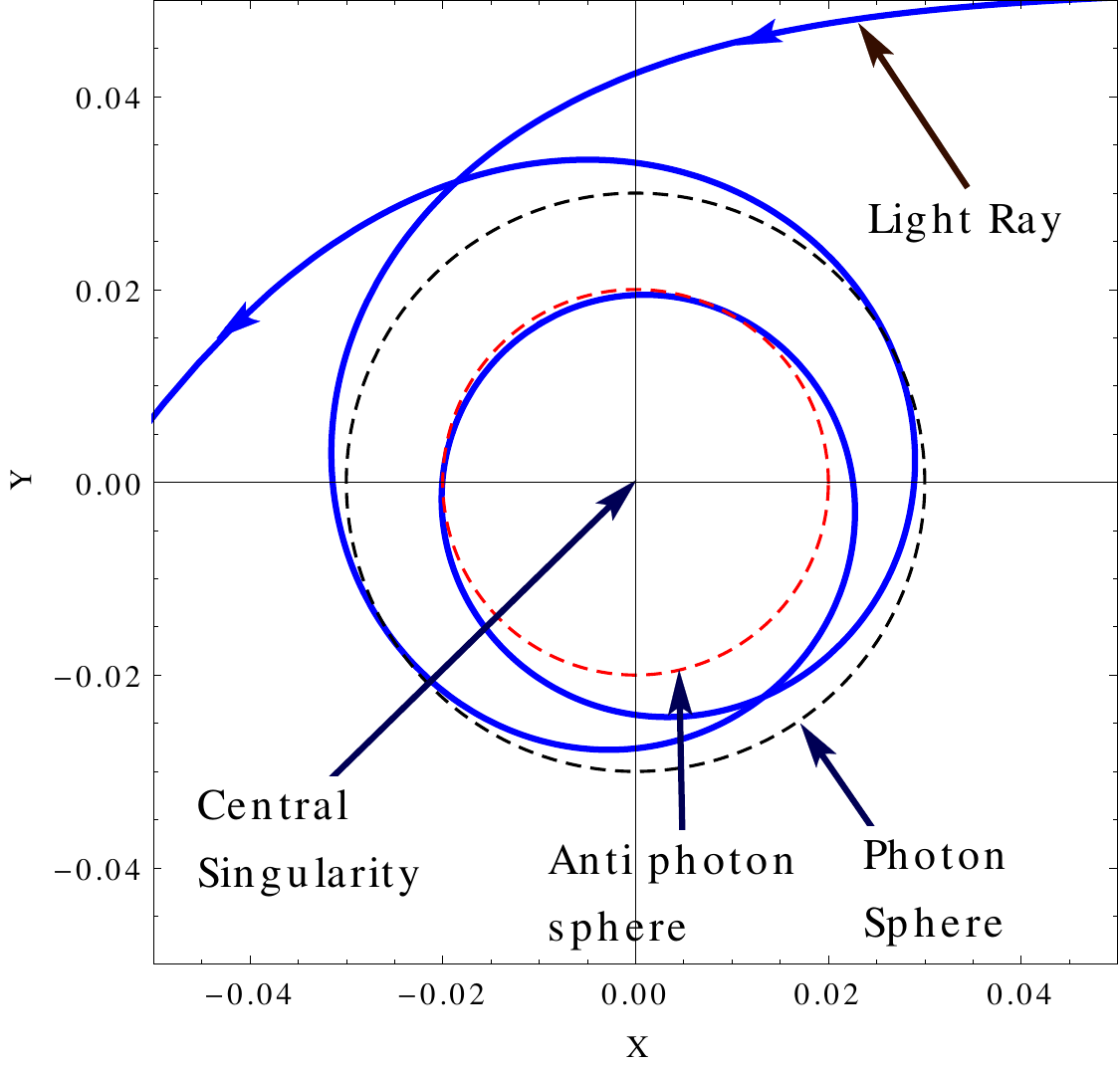}
\caption{A typical photon trajectory with $2M<R_m<3M$ that experiences an anti-photon sphere. 
Here we have taken $M=0.01$, $\alpha = 1.2$, $b = 0.5$.}
\label{traj1}
\end{minipage}
\hspace{0.4cm}
\begin{minipage}[b]{0.5\linewidth}
\includegraphics[width=3.2in,height=3.0in]{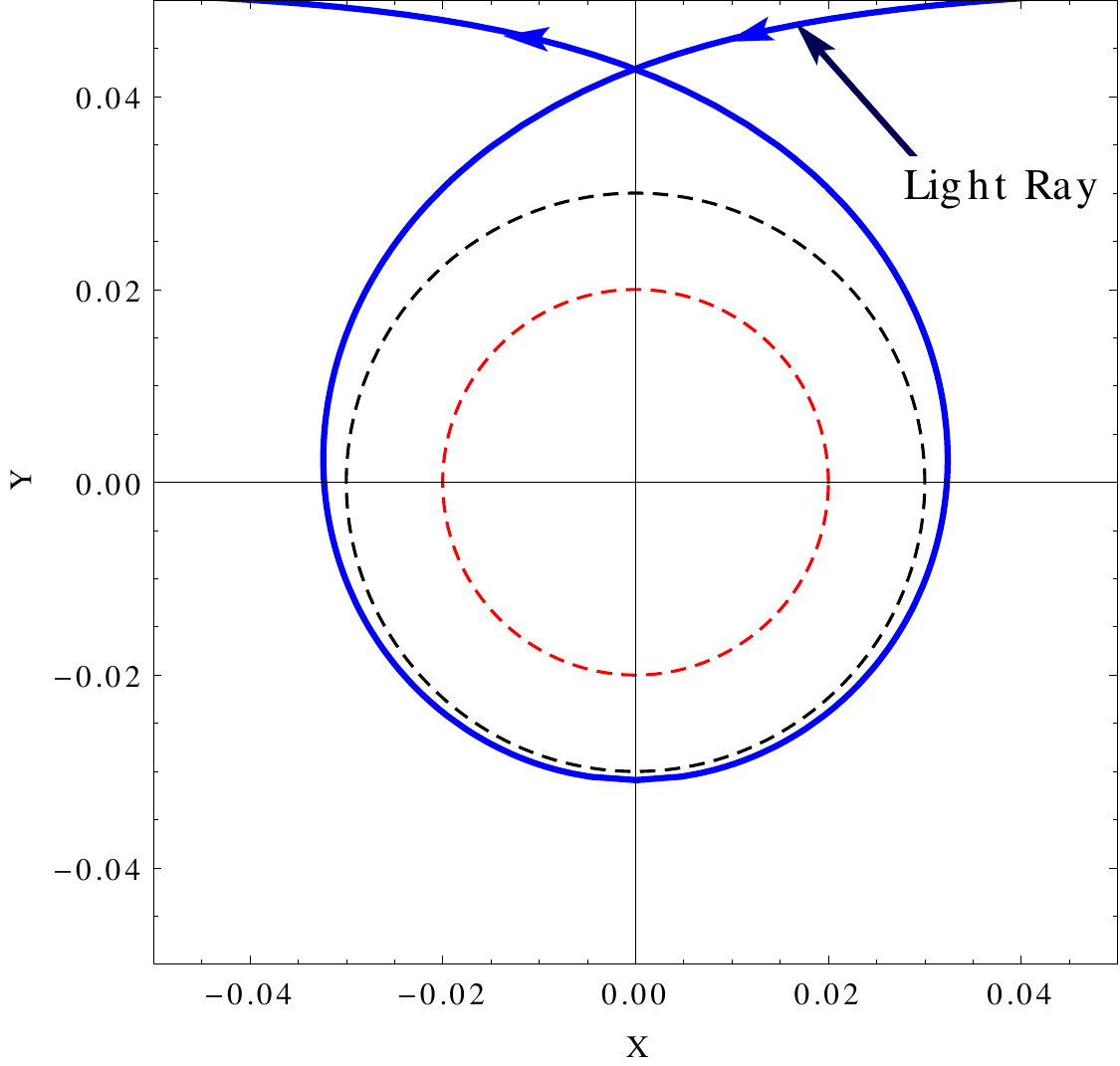}
\caption{A typical photon trajectory with $2M<R_m<3M$ that does not experience an anti-photon sphere. 
Here we have taken $M=0.01$, $\alpha = 1.2$, $b = 0.5$.}
\label{traj2}
\end{minipage}
\end{figure}

A few distinct situations might arise in such cases where the light rays (possibly coming from a distant source) 
pass close to the photon sphere, i.e undergo strong lensing. In the first scenario, with $2M<R_m<3M$, 
photons are trapped at the location of the
external Schwarzschild photon sphere at $q=3M$. Under the influence of a small perturbation, these might escape to an
observer at infinity. In this case, there will be relativistic images (Einstein rings) that will be entirely identical to a Schwarzschild
black hole. However, within this framework, 
a second more interesting case might occur, when the rays trapped at the location of the photon sphere 
at $q=3M$, might proceed to a lower radius with a small perturbation, then reflect from the potential barrier inside and
finally escape to an observer at infinity after some further rotations at the photon sphere. The situation has been recently addressed
in \cite{Tapo} and the final result is that this case is qualitatively different from strong lensing
by black holes \cite{Bozza1}. Also, for $R_m > 3M$, 
when there are no extrema of the effective potential, there are no Einstein rings,
and light suffers ordinary deflection from a potential barrier. We will analyze these situations further below. 

\subsection{Lensing from the singular solution}

To elaborate on the points that we have just mentioned, in fig.(\ref{traj1}), we numerically integrate the null geodesic equation
in the background of the metric of eq.(\ref{dsfinal}), and plot a typical photon trajectory, with $M=0.01$, $\alpha = 1.2$ 
and $b = 0.5$. $R_m$ is
determined to be $0.028$ here from eq.(\ref{mass}), so that $2M<R_m<3M$, also, eq.(\ref{rFinal}) dictates that
in order to be in the physical region, one requires that $q<0.037$, a condition satisfied here. 
In this case, the photon hits the internal potential 
barrier (see fig.(\ref{fig1})) and we have chosen
a turning point close to where the height of the effective potential is the same as that of the photon sphere (the point $q_1$
in fig.(\ref{fig1})). As discussed, 
the photon suffers multiple turnings before escaping to an observer at infinity. Fig.(\ref{traj2}) shows a typical photon trajectory for
the same set of parameters but with a different turning point, so that the photon does not experience the effect of the 
anti photon sphere, and is reflected from the outer Schwarzschild photon sphere. 

To quantify these cases, as before, we will set the polar angle $\theta = \pi/2$ and 
recall \cite{Weinberg} that in gravitational lensing, the deflection angle $\alpha(q_{0})$ of the light coming from a distant source, 
where $q_0$ is the turning point of light at the distance of closest approach from the lens is obtained as
\begin{equation}\label{defl}
\alpha(q_{0})=I(q_{0})-\pi,
\end{equation}
where
\begin{equation}\label{eq:I1}
I(q_{0})\equiv 2\int^{\infty}_{q_{0}}\frac{dq}{\sqrt{\frac{R(q)g_{\phi\phi}(q)}{g_{qq}(q)}}}~,~
R(q)= \left(\frac{g^0_{tt}g_{\phi\phi}}{g_{tt}g^0_{\phi\phi}}-1 \right)~,
\end{equation}
where a superscript $0$ indicates that the corresponding quantity is evaluated at the turning point. 
\begin{figure}[h!]
\begin{minipage}[b]{0.5\linewidth}
\centering
\includegraphics[width=3.2in,height=3.0in]{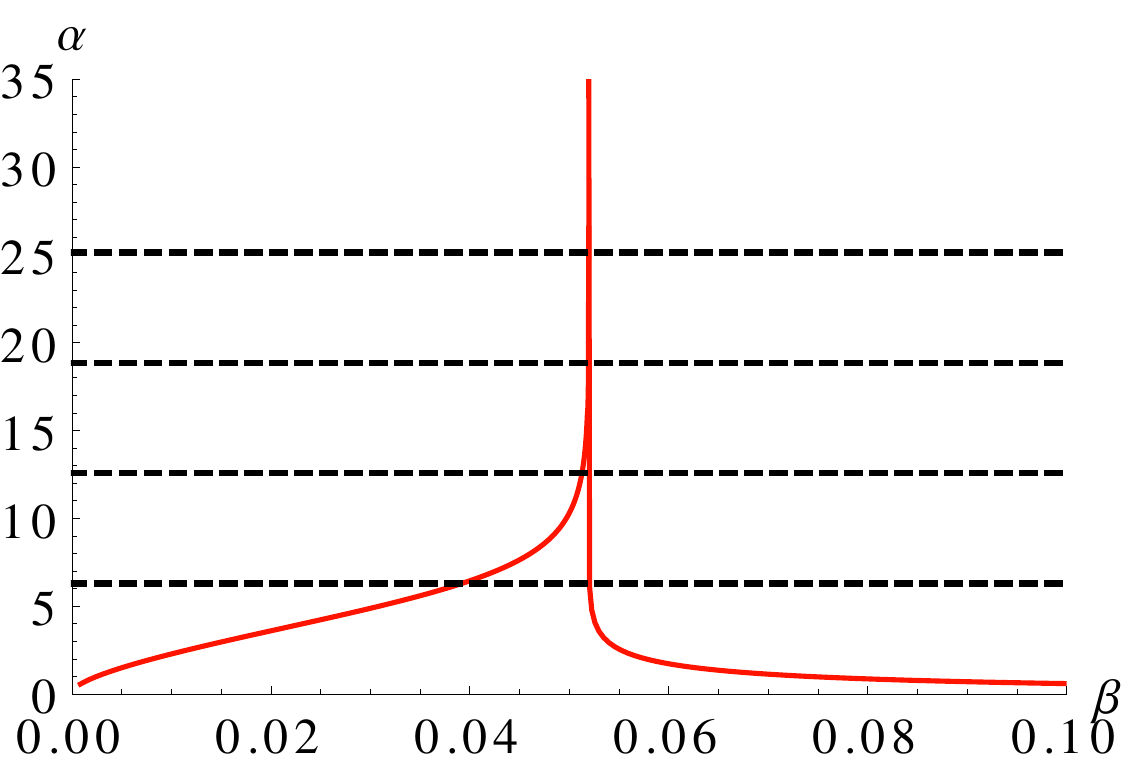}
\caption{Deflection angle as a function of the impact parameter for $M=0.1$, $\alpha = b = 0.5$, see text.}
\label{def1}
\end{minipage}
\hspace{0.4cm}
\begin{minipage}[b]{0.5\linewidth}
\includegraphics[width=3.2in,height=3.0in]{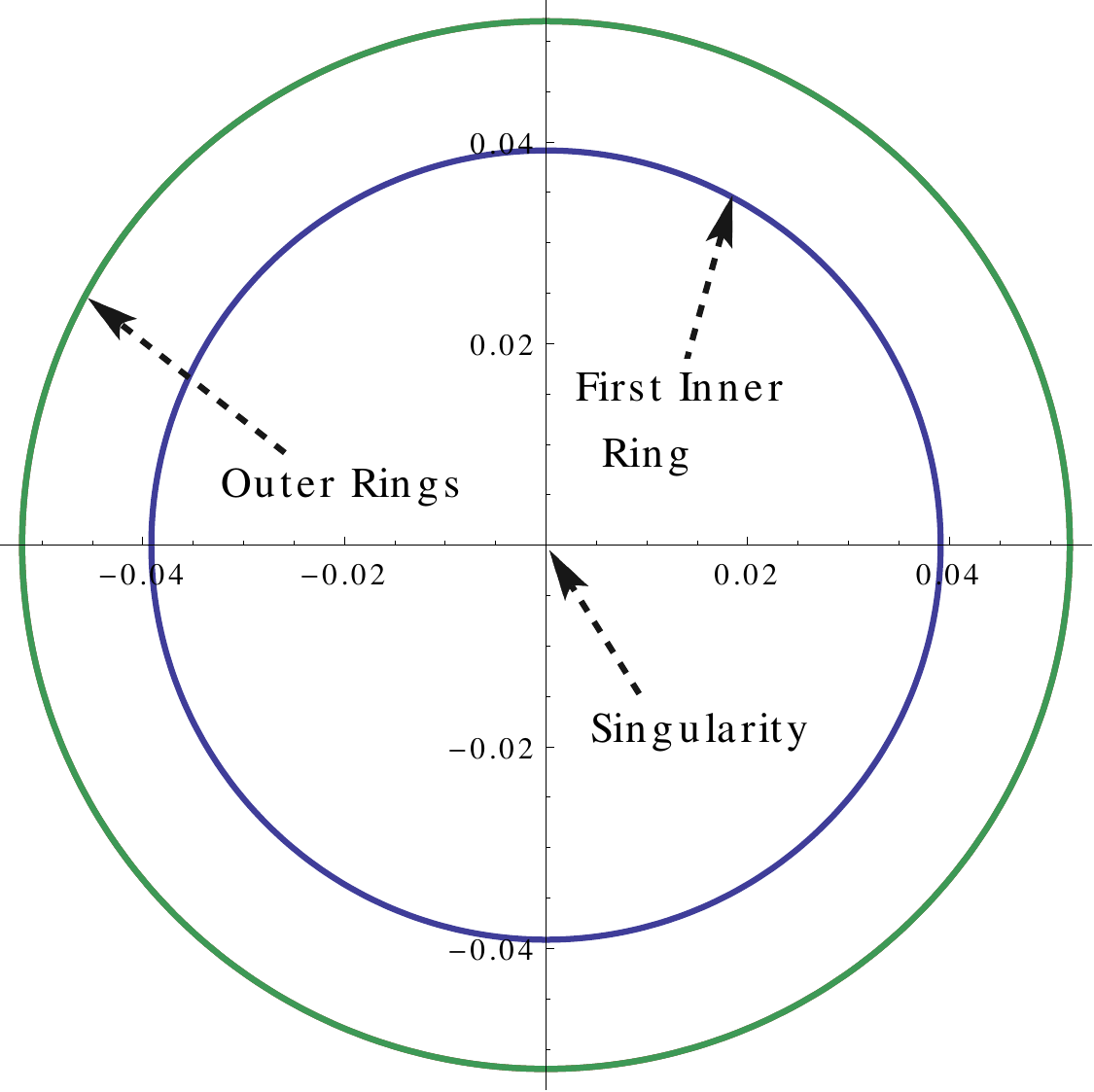}
\caption{Einstein rings in the observer's sky for $M=0.1$, $\alpha = b = 0.5$, see text.}
\label{def2}
\end{minipage}
\end{figure}

We may evaluate the deflection angle numerically from the metric of eq.(\ref{dsfinal}). 
In fig.(\ref{def1}), we take $M=0.01$, $\alpha =1.2$, $b = 0.5$, and
plot the deflection angle as a function of the impact parameter of the light rays, $\beta = L_{ph}/E_{ph}$, which translates to the
well known result that for a spherically symmetric metric, $\beta(q_0) = \sqrt{q_0^2/(-g_{tt}(q_0))}$. This angle diverges
at the location of the photon sphere, and in this figure,
we have shown by dashed horizontal lines $\alpha = 2\pi, 4\pi, 6\pi \cdots$, where the Einstein rings are observed
corresponding to the photon experiencing multiple windings, assuming that the source (at infinity),
the lens and the observer (at infinity) are aligned. Note from fig.(\ref{def1}) that the first Einstein ring inside
the photon sphere occur at a value of the impact parameter that is distinctly lower than {\it all} other rings. The structure of 
the Einstein rings in the observer's sky for our case 
is shown in fig.(\ref{def2}). Here, the innermost image is the first Einstein ring, which is distinct from
all others that are represented by the outer blue circle. This is a novel structure and we are not aware of such a system
of Einstein rings in the present literature. 

Clearly, the anti-photon sphere plays a crucial role in this analysis, and to glean insight into realistic situations, we will
now use some results from \cite{Tapo} where this was investigated analytically in the case of strong lensing, i.e where 
photons pass close to the photon sphere. We will assume such a limit, and simply borrow the result that 
it can be shown (for details see \cite{Tapo}), that the deflection angle of light experiencing 
strong lensing, in the presence of an anti-photon sphere and forming images inside the photon sphere 
is given in terms of its impact parameter $\beta$, with a critical value
$\beta_c$ at the photon sphere as
\begin{equation}
\alpha(\beta)=-\bar{a}\log \left( \frac{\beta_c^2}{\beta^2}-1 \right) +\bar{\beta} +\mathcal{O}((\beta_c^2-\beta^2)\log(\beta_c^2-\beta^2)),
\label{eq:strong_alpha_2}
\end{equation}
where
\begin{equation}
\bar{a}=2\sqrt{\frac{2g^m_{qq} g^m_{tt}}{g_{\phi\phi}^{'' m} g^m_{tt}-g^m_{\phi\phi} g_{tt}^{'' m}}},
\label{eq:strong_abar_2}
\end{equation}
\begin{equation}
\bar{\beta}=\bar{a}\log \left[2q_m^2\left(\frac{g_{\phi\phi}^{'' m}}{g^m_{\phi\phi}}-\frac{g_{tt}^{'' m}}
{g^m_{tt}}\right)\left(\frac{q_m}{q_1}-1\right)\right] +I_R(q_1)-\pi.
\label{eq:strong_bbar_2}
\end{equation}
Here, $I_R(q_1)$ is a regulated integral that is free of the divergences in the integral appearing in 
eq.(\ref{defl}), and a superscript $m$ on a quantity indicates that it is evaluated at the photon sphere whose location is $q_m$, 
and $q_1$ is a radius where the height of the effective potential equals that at the photon sphere, for example 
the dashed vertical line in fig.(\ref{fig1}). Eqs.(\ref{eq:strong_alpha_2}), (\ref{eq:strong_abar_2}) and (\ref{eq:strong_bbar_2})
are valid in the limit $q_0 \to q_1$, i.e the turning point of photons inside the photon sphere happens close to the point
where the height of the effective potential equals that at the photon sphere.  

Observables in the outer image system are defined in the standard way following \cite{Bozza1}. We start with the lens equation
for gravitational lensing in the strong field limit, given by 
\begin{equation}\label{lens}
\delta=\theta-\frac{D_{LS}}{D_{OS}}\Delta\alpha_n~.
\end{equation}
Here, $\delta$ is the angular 
separation between the lens and the source, $\theta$ is the angular separation between the lens and the relativistic image 
formed, and $\Delta\alpha_n=\alpha(\theta)-2\pi n$ denotes the offset of the deflection angle. 
Also, $D_{LS}$ denotes the lens to source distance, $D_{OS}$ is the observer to source distance, and
$D_{OS}=D_{OL}+D_{LS}$, $D_{OL}$ being the distance between the lens and the observer. 
This is obtained after subtracting all 
the winding that the photon undergoes (encoded in the integer $n$). 

With this in mind, the observables in strong lensing for outer images are as follows \cite{Bozza1} 
\begin{equation}
\theta_n=\frac{\beta_c}{D_{OL}}(1+e_n)~,~\mu_n = \frac{\beta_c^2 D_{OS}e_n(1+e_n)}{\bar{a}D_{OL}^2 D_{LS}}~,~
s_1=|\theta_1-\theta_\infty|~, ~ {\rm where}~e_n=e^{\frac{\bar{\beta}-2n\pi}{\bar{a}}}~,~
\end{equation}
For the inner image system, the following quantities can be defined by analogy \cite{Tapo} :
\begin{eqnarray}
&~&\theta_{-n}=\frac{\beta_c}{D_{OL}}\frac{1}{\sqrt{1+e_{-n}}}~,~
\mu_{-n} = -\frac{\beta_c^2 D_{OS}}{2\bar{a}\beta D_{OL}^2 D_{LS}}\frac{e_{-n}}{(1+e_{-n})^2}~,~
s_{-n}=|\theta_{-n}-\theta_{-(n+1)}|~,~\nonumber\\
&~& {\rm where}~e_{-n}=e^{\frac{\bar{\beta}-2n\pi}{\bar{a}}}.
\end{eqnarray}
With these ingredients, we present our numerical results for these variables in Table (\ref{Table1}). For comparison, we have
presented the results for the Schwarzschild geometry, where there are no inner images. 
\begin{table}[h!]
\centering
\caption{The angles are in micro arc sec. Here, $\alpha = 1.2$, $M=10^7 M_{\odot}$, $D_{OL}=D_{LS}=8$ Kpc.}
 \begin{tabular}{| c | c | c | c | c |} 
 \hline\hline
  & Schwarzschild & \multicolumn{3}{c|}{Metric of eq.(\ref{dsfinal})} \\
  & black hole & \multicolumn{3}{c|}{$~$}  \\
  & & b=0.46 & $b=0.48$ & $b=0.50 $ \\
 \hline
  $\theta_1$ & 0.6447 & 0.6447 & 0.6447 & 0.6447 \\
  $\theta_{\infty}$ & 0.6439 & 0.6439 & 0.6439 & 0.6439 \\
  $\theta_{-3}$ & --- &0.6428 & 0.6431 & 0.6434 \\
  $\theta_{-2}$ & --- & 0.6213 & 0.6271 & 0.6339  \\
  $\theta_{-1}$ & --- & 0.3911 & 0.4289 & 0.4890 \\
  $\mu_1\times 10^{25}$ & 2.7983 & 2.7983 & 2.7983 & 2.7893 \\
  $\mu_{-3}\times 10^{25}$ & --- & $-1.7717$ & $-1.3011$ & $-0.7626$  \\
  $\mu_{-2}\times 10^{25}$ & --- & $-35.7771$ & $-27.2190$ & $-16.6255$  \\
  $\mu_{-1}\times 10^{25}$ & --- & $-129.8900$ & $-137.8040$ & $-136.2740$  \\
  $s_1$ & 0.0081 & 0.0008 & 0.0008 & 0.0008  \\
  $s_{-3}$ & --- & 0.0009 & 0.0007 & 0.0004  \\
  $s_{-2}$ & --- & 0.0215 & 0.0160 & 0.0095  \\
  $s_{-1}$ & --- & 0.2302 & 0.1982 & 0.1449 \\
 \hline\hline
 \end{tabular}
\label{Table1}
\end{table}
For illustration, and to provide a realistic scenario, 
we have taken the central mass as $10^7M_{\odot}$. The observer - lens and source - lens distances
are chosen as to $8~{\rm Kpc}$. These are typical values for the black hole Sgr A* at our galactic center and are
chosen to provide a quantitative contrast with the black hole background.
The angular separation $\delta$ between the lens and the source in eq.(\ref{lens})
is taken as $5$ degrees. In this table, by $\theta_{\infty}$ we mean the $1000$th image outside the photon sphere. 

Table (\ref{Table1}) indicates that there are substantial differences between the images formed outside the photon sphere
and those formed inside it, for strong gravitational lensing.\footnote{We note here that there might be a small error in the exact
value of $\theta_{-1}$, and the value might be slightly smaller than the one we show here. This is because in this case, the 
limit $q_0 \to q_1$ mentioned before might not be strictly true, see discussion after 
eq.(\ref{eq:I1}). We have not been careful about this here, as the value reproduced here serves to illustrate our point.} 
As mentioned, the images outside the photon sphere for our naked singularity are formally identical to the Schwarzschild case, and
one cannot distinguish between the two, only on this basis. Also, it is seen from Table (\ref{Table1})
that as indicated before, except for the innermost Einstein
ring, all the others form at almost the same location in the observer's sky. However, 
the magnification of the first image inside the photon sphere is almost two orders of magnitude
higher than the first image outside it. Similarly, the angular separation between the first and second images inside the photon
sphere can be almost three orders of magnitude higher than that between the first and the $1000$th image outside the photon
sphere.  As was noted in \cite{Tapo}, these are typical features in ultra-compact horizon-less objects. 

\subsection{Accretion disks around the singular solution}

We will now briefly address the issue of thin accretion disks around the naked singularity solution of eq.(\ref{dsfinal}),
following the results derived at the beginning of this section. As we have pointed out before, we consider massive particles
in accretion discs around the singular region with the assumption that these do not interact with the matter seeding the metric,
but only move under the influence of geometry. Indeed, such assumptions are well known in the literature (see, e.g
\cite{Joshi:2011zm}, \cite{Kovacs:2010xm}, \cite{JMN2}). 

The quantity of interest here will be the matching radius $R_m$. If we match our solution of eq.(\ref{dsfinal}) with
an external Schwarzschild solution with $2M<R_m<3M$, then it can be checked that there are no values of $(\alpha, b, M)$
for which $q_{max} > R_m$, with $q_{\rm max}$ defined in eq.(\ref{qmax}). 
These were exemplified in the previous subsection (see discussion after eq.(\ref{ps})) as
cases where an anti-photon sphere exists. We therefore conclude that in these cases, stable closed circular orbits will
exist up to an OSCO with radius $q_{max} < R_m$ in the internal metric, after which such stable circular orbits will exist 
beyond the innermost stable circular orbit (ISCO) of the external Schwarzschild solution. Hence, 
whenever a photon sphere exists, the accretion disk
will consist of two disjoint regions consisting of $(0<q<q_{max} \cup R_m<q<\infty)$.

The situation changes qualitatively when we consider matching with an external Schwarzschild solution 
with $R_m > 3M$, i.e there is no photon sphere (see fig.(\ref{fig1})). Here we find that there exist possible
values of  $(\alpha, b, M)$ for which $q_{max} > R_m$. Analytic expressions for the constraints on the variables
is mathematically cumbersome and not very illustrative. Here again there is a possibility of disjoint accretion disks, 
when $R_m < 6M$, which are qualitatively similar to the case discussed in the previous paragraph. 
We will avoid these, and present an example in which we will demand that $R_m > 6M$ with $q_{max} > R_m$, 
so that we will have a continuous accretion disk. 

To study basic properties of thin accretion disks in our model, we follow the pioneering work of Novikov and Thorne 
\cite{Thorne:1974ve} and Page and Thorne \cite{Page:1974he} who generalised the Newtonian approach known in
the literature as the Shakura-Sunyaev model \cite{ShakuraSunyaev}. This is an extremely well researched topic and
we will simply state the main result, that from local frame of the accreting fluid, the flux radiated by the disk 
is given as 
\begin{eqnarray}\label{flux}
f(q) = - {\dot m\over \sqrt{-g}} {\omega,_q\over (E-w L)^2}\int_0^q (E-w L) L,_{{\hat q}} d{\hat q}\, ,
\end{eqnarray}
where $E$, $L$ and $\omega$ are obtained from eqs.(\ref{el}) and (\ref{omnew}). 
Here $\dot m$ is mass accretion rate of the metric, which is usually taken to be a constant, and $\sqrt{-g}$ is the 
determinant of the 3-metric with $\theta = \pi/2$. 
Further, for steady state thin disk models, when the accreting matter is in thermodynamic equilibrium, the energy
flux of eq.(\ref{flux}) can be used to define a temperature of the disk $T$ via the Stefan-Boltzmann law, and this in turn can
be used to compute the luminosity at spatial infinity, assuming that the radiation emitted by the disk has a black body
spectrum. The final result is (see e.g \cite{Torres},\cite{Kovacs:2010xm})
\begin{equation}\label{lumin}
L(\nu) = \int_{q_i}^{\infty}\frac{\nu_e^3 qdq}{{\rm exp}(\nu_e/T) - 1}~,
\end{equation}
where we will set the Stephan-Boltzmann constant, the Boltzmann's constant and the mass accretion rate to
unity, as we will be interested in a qualitative classification
of the radiation spectrum compared to the black hole case. Note that here $\nu_e$ is the emitted frequency, which is
related to the observed frequency $\nu$ by the redshift factor $z$ as, $\nu_e = (1+z)\nu$. The redshift factor
is defined as \cite{Luminet}
\begin{equation}
1+z = \frac{1}{\left[-g_{tt} - \omega^2g_{\phi\phi}\right]^{\frac{1}{2}}}~,
\end{equation}
where we have set the inclination angle of the disk to zero and assumed the radiation to be isotropic. 

\begin{figure}[h!]
\begin{minipage}[b]{0.5\linewidth}
\centering
\includegraphics[width=3.2in,height=2.8in]{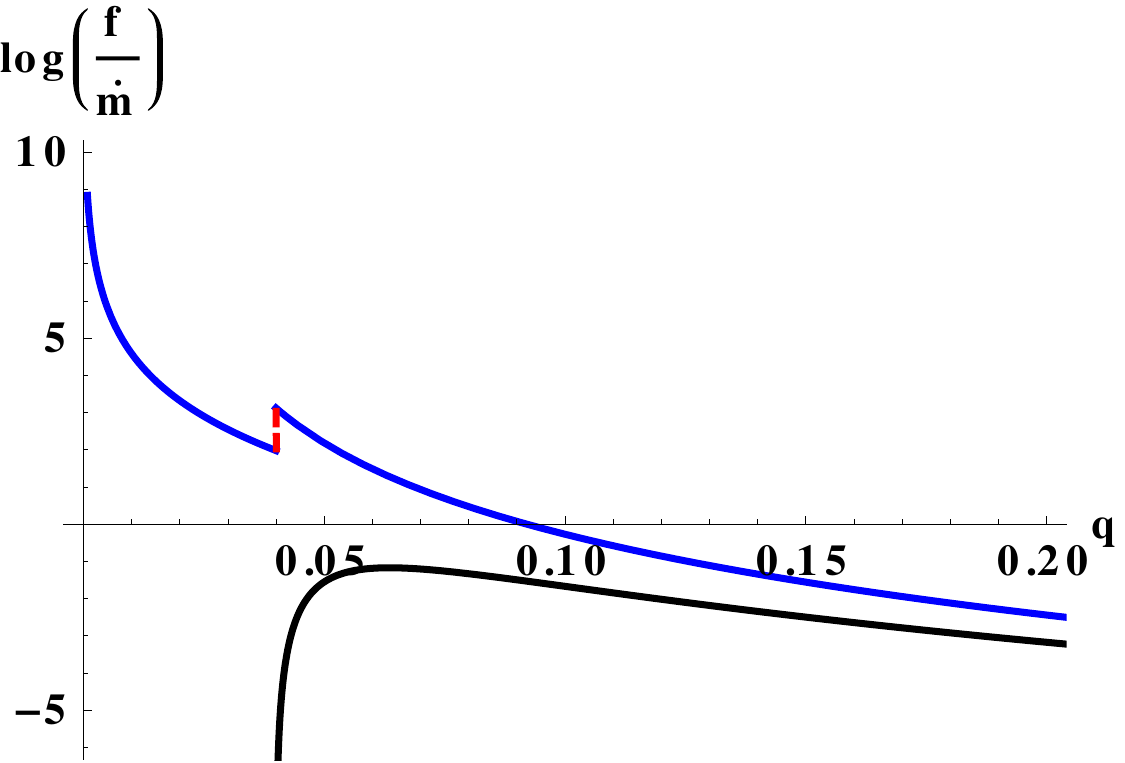}
\caption{$f(q)/{\dot m}$ for $R_m=0.04$ and $\alpha=1.2, b=0.95$ corresponding to solid blue lines. 
The red vertical line represents the jump in flux. The black line corresponds to the Schwarzschild black hole.}
\label{fig2}
\end{minipage}
\hspace{0.4cm}
\begin{minipage}[b]{0.5\linewidth}
\includegraphics[width=3.2in,height=2.8in]{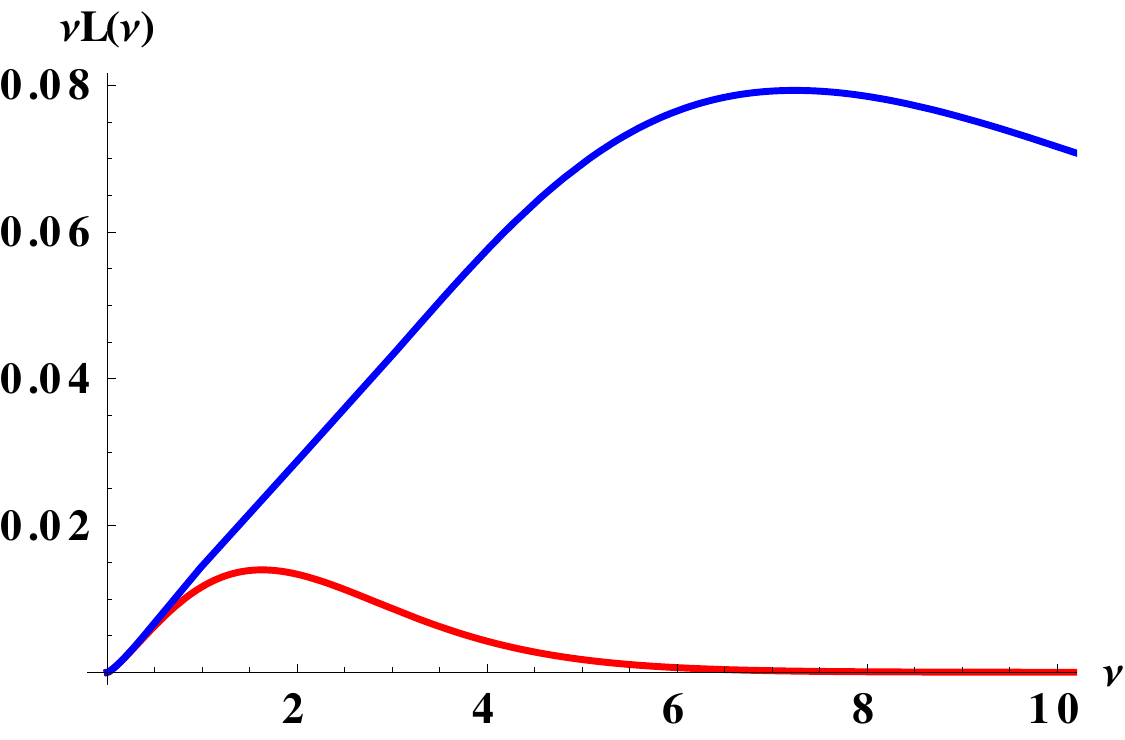}
\caption{The solid blue line is $\nu L(\nu)$ of the accretion disk of eq.(\ref{dsfinal}) 
as a function of the observed frequency for $R_m=0.04$, with $\alpha=1.2$, $b=0.95$. The solid red line 
 is the corresponding quantity for a black hole of the same mass. }
\label{fig3}
\end{minipage}
\end{figure}

As an illustrative example, we will choose here $\alpha = 1.2$, $b=0.95$ and $R_m = 0.04$, so that we are in the physical 
range defined by eq.(\ref{con3}), and also the mass, calculated to be $M=0.00665$ is such that $R_m>6M$. 
In figure (\ref{fig2}), we show the flux per unit mass accretion rate, $\log(f(q)/{\dot m})$ by the solid blue curve.
In the figure, the dotted red vertical line represents a jump in the flux, which 
arises due to the discontinuity of the derivative of the angular speed of eq.(\ref{omnew}) across the matching
surface. The flux per unit mass accretion rate for the Schwarzschild black hole with the same mass is shown
by the solid black line. 

Finally, in fig.(\ref{fig3}), we plot the quantity $\nu L(\nu)$ of eq.(\ref{lumin}) as a function of the frequency $\nu$ for
the same range of parameters discussed in the previous paragraph, where, as mentioned before, all constants
appearing in eq.(\ref{lumin}) have been set to unity. 
The solid blue line shows this quantity for the
naked singularity of eq.(\ref{dsfinal}). The solid red line is the black hole result with the mass determined 
from eq.(\ref{mass}). The qualitative differences in the nature of these graphs are clearly noticeable, namely that $\nu L(\nu)$
peaks at a lower frequency for the black hole, compared to the naked singularity case, and drops off to small values
in the former case much faster. 

Before closing this section, we briefly comment on the radiative efficiency $\eta$ of the naked singularity of eq.(\ref{dsfinal}). In
the Novikov-Thorne model, the standard definition of this efficiency is (see e.g. \cite{Bambi}) $\eta = 1 - E_{ISCO}$
where $E_{ISCO}$ is the energy per unit rest mass of a particle at the innermost stable circular orbit. In our case, since
stable circular orbits are possible all the way up to $q=0$, we have from the first equation in eq.(\ref{le}), 
\begin{equation}
\eta = \left(1 - \sqrt{-g_{tt}}\right)\bigr |_{q\to 0} = 1- \left[1-\frac{1}{b}
\left(\frac{R_m}{b}\right)^{\frac{2-\alpha}{1+\alpha}}\right]^{\frac{3}{4-2\alpha}}
\end{equation}
We illustrate this by choosing a few typical values that we have used in this section. For $M=0.01$,
$\alpha = 1.2$, $b=0.5$, we obtain a radiative efficiency of $\sim 90\%$, while for $R_m = 0.04$, $\alpha = 1.2$
and $b = 0.95$, we obtain a radiative efficiency of $\sim 95\%$. 
This of course means that our solution might be affected by the infalling matter. 
In this context, note that for the Schwarzschild black hole, the radiative efficiency
is $\sim 6\%$ \cite{Bambi} whereas for the JMN naked singularity, this efficiency is $100\%$  \cite{Joshi:2011zm}. 
Clearly, our solution then is qualitatively different from the JNM singularity in this aspect. On physical grounds, we do expect
accreting matter to cause a change in the physical properties of the central singularity. How this happens in our case
is an involved question, which we hope to return to in the future.

\section{Conclusions and Discussions}
\label{concl}
It is well known by now that with suitable conditions, gravitational collapse from a set of regular initial data 
might be made to halt, with the resulting meta-stable state being sufficiently long-lived to warrant physicality. 
These nonetheless have a central singularity, which is visible to an external observer at spatial infinity, and
thus present examples of naked singularities. 
In the present paper we have studied such a collapse process, and used the junction conditions to match 
an expanding FLRW metric with a generic contracting solution of Einstein's equations, on a space-like
hyper-surface, and studied some observational aspects of the resulting two-parameter family of static 
naked singularity backgrounds. Although such a static background will be reached in very large co-moving time, 
we have seen that there exists a sufficiently large neighborhood near the
equilibrium co-moving time in which the equilibrium configuration is approximately static.
As we have seen, our equilibrium metric is very different from the one obtained by a similar
analysis carried out in \cite{Joshi:2011zm}. Our solution is thus novel, and adds to the literature
on naked singularities. As mentioned in the introduction, our collapse model is distinct from the conventional
structure formation scenarios via virialization, and can be understood as a toy model to study the formation
of a naked singularity.

As we have mentioned, currently there is a lot of interest in the observational distinctions between
black holes and naked singularities. In this spirit, we have modelled the central singularity as a typical
$10^7M_{\odot}$ object, and studied aspects of gravitational 
lensing. Accretion disk properties of our model have also been considered, and both these have been
contrasted with the corresponding black hole backgrounds with similar mass.
As for the former, we have shown that interesting lensing phenomena can arise, with a possible anti-photon sphere
playing a crucial role, as was the case in the analysis of \cite{Tapo}. The Einstein ring system here is novel, and different
from the ones known in the literature. The behavior of accretion disks here were also shown to 
be distinct from black hole backgrounds
(thin disk properties of some other naked singularity models have been studied in \cite{Kovacs:2010xm}). 
Indeed, with the advent of the Event Horizon Telescope era, it is important and interesting to document such results
as they often mimic black hole phenomena. There already exists a large body of literature on the observational 
signature of naked singularities, and our work adds to this, with a novel background metric. 

We have further established that although the solution presented here does not have an obvious equation of state,
it can nonetheless be modeled by a two-fluid scenario. In this picture, our energy momentum tensor consists of 
two non-interacting perfect fluids, one of which might be dust. In the case where the other component consists of
stiff matter, the energy density of the composite fluid is carried over by the dust, and the pressure anisotropy acts
as the energy density of stiff matter. 

An obvious question that might arise in this context is regarding the assumption of vanishing radial pressure in our model.
Arguably, this was an assumption that simplified some of the computations (a generic model allowing 
radial pressure is substantially complicated, and we  hope to report on this in future). As we have amply
pointed out, this assumption is not uncommon in the literature, with the most famous example being the 
Einstein cluster, and the interior Schwarzschild solution due to Florides. Our result can be thought
of as a generalization of the regular Floride's interior metric to situations with a central singularity. Note that in 
most astrophysical models of galactic centers, the density profile shows a divergence at the origin and has
to be cut-off at a small radius. Whether quantum corrections provide such an escape route in GR is 
currently not known. As of now, we present our solution of eq.(\ref{dsfinal}) as a specific GR example, without any
claims of identifying these with more realistic galactic models. 

Before ending, we point out two feature of this analysis that might be interesting to explore further. First, as pointed
at the end of the previous subsection, it might be interesting to further consider accretion properties of our model, as we
have seen that this is affected by infalling matter. Secondly, 
we did not consider lensing effects in our study on accretion disks. Indeed, as we have mentioned, the photon effective potential
behaves in a very different way here, compared to black holes. Here, in the case of disjoint accretion disks, 
all radiated energy from the inner disk will not reach 
an observer at infinity, but might be trapped by the effective potential , i.e turn back towards the singularity \cite{PhenRev}.
Such phenomena might assume significance in the physics of accretion disks around naked singularities, and should
be further investigated. 
\begin{center}
{\bf Acknowledgements}
\end{center}

TS would like to thank Rajibul Shaikh, Pritam Banerjee and Suvankar Paul for useful discussions and comments. 

\bibliographystyle{mystyle}
\bibliography{collapse-ref.bib}
\end{document}